\documentclass{article}
\usepackage{amssymb,amsfonts,amsmath,stmaryrd}
\usepackage{indentfirst}
\usepackage{enumerate,float}
\usepackage{color}
\usepackage{tikz}
\usetikzlibrary{arrows,snakes,backgrounds}
\usepackage{ytableau}
\usepackage{physics}
\usepackage{comment}
\usepackage{setspace}
\usepackage[hidelinks,colorlinks=true,unicode]{hyperref}
\hypersetup{linkcolor=orange, citecolor=orange, filecolor=orange, urlcolor=orange}
\usepackage{url}
\usepackage[natbib=true, style=numeric-comp, sorting=none, maxbibnames=10]{biblatex}
\def\be{\begin{eqnarray}}
\def\ee{\end{eqnarray}}

\def\p{\partial}
\def\tr{{\rm tr}\,}

\def\yt{\scriptsize \text }

\definecolor{red}{rgb}{1,0,0}
\definecolor{orange}{rgb}{1,0.5,0}
\definecolor{violet}{rgb}{0.7,0,1}

\hoffset= - 1.0in         

\begin{document}
\title{{\Large {\bf 3-Schurs  from explicit representation of  Yangian  $Y(\hat{\mathfrak{gl}}_1)$. \\
Levels 1-5}
\vspace{.1cm}}
\author{
{\bf A.~Morozov $^{a,b,c,d}$},
{\bf  N.~Tselousov $^{a,b,d}$} \date{ }}
}
\maketitle

\vspace{-5.5cm}

\begin{center}
\hfill MIPT/TH-12/23\\
\hfill IITP/TH-11/23\\
\hfill ITEP/TH-15/23
\end{center}

\vspace{3cm}

\begin{center}
\begin{small}
$^a$ {\it MIPT, 141701, Dolgoprudny, Russia}\\
$^b$ {\it NRC “Kurchatov Institute”, 123182, Moscow, Russia}\\
$^c$ {\it IITP RAS, 127051, Moscow, Russia}\\
$^d$ {\it ITEP, Moscow, Russia}\\
\end{small}
\end{center}

\vspace{0.5cm}

 
\begin{abstract}
{\footnotesize
We suggest an ansatz for representation of affine Yangian $Y(\hat{ \mathfrak{gl}}_1)$ by differential operators
in the triangular set of time-variables ${\bf P}_{a,i}$ with $1\leqslant i\leqslant a$, which saturates the MacMahon formula for the number of $3d$ Young diagrams/plane partitions. In this approach the 3-Schur polynomials are defined
as the common eigenfunctions of an infinite set of commuting "cut-and-join" generators $\psi_n$ of the Yangian.  We manage to push this tedious program through to the 3-Schur polynomials of level 5, and this provides a rather big sample set, which can be now investigated by other methods.

}
\end{abstract}


\section{Introduction}

\subsection{The purpose of the paper}

The simplest technical definition 
of Schur functions $S_R$ \cite{Macdonald} describes them as common eigenfunctions
of commuting system of "cut-and-join" operators $\hat W_{\Delta}$ \cite{MMN}, that are part of $W_{1 + \infty}$ algebra \cite{PSR,AFMO}:
\be
\hat W_{\Delta} S_R = \lambda_{\Delta,R} S_R
\ee
where $R=[R_1, \ldots, R_{l(R)}]$ and $\Delta = [\Delta_1, \ldots, \Delta_{l(\Delta)}]$ are Young diagrams (integer partitions) of equal size
and eigenvalues $\lambda_{\Delta,R}$ -- are the characters of symmetric group $s_N$. The operators are defined in the following way
\begin{equation}
    \hat W_\Delta = \frac{1}{z_{\Delta}}\ :  \prod_{k=1}^{l(\Delta)} \tr \hat D^{\Delta_k} : \hspace{15mm} \hat D_{a,b} = \sum_{c=1}^{N} X_{a,c} \frac{\p}{\p X_{b,c}}
\end{equation}
where $z_{\Delta}$ is a standard factor \cite{MMN} and $X_{a,b}$ is a $N \times N$ matrix. Introducing {\it time}-variables $p_a$ and performing the Miwa transformation $p_a=\tr X^a$ the simplest cut-and-join operator takes the following form:
\begin{equation}
\ytableausetup{boxsize = 0.4em}
    \hat W_{\, \begin{ytableau}
        \ & \ 
    \end{ytableau}} = \frac{1}{2} \sum_{a,b = 1}^{\infty} \left(ab \cdot p_{a+b}\frac{\p^2}{\p p_a \p p_b} + (a+b) \cdot p_ap_b\frac{\p}{\p p_{a+b}}\right)
\end{equation}
\begin{equation}
    \lambda_{\, \begin{ytableau}
        \ & \ 
    \end{ytableau}, R} = \sum_{\Box \in R} \, \left( x_2(\Box) - x_1(\Box) \right)
\end{equation}

The purpose of this paper is to define in the same way the $3$-Schur functions
${\bf S}_\pi$, where $\pi$ are the $3d$ Young diagrams (plane partitions) and the time variables are two-parametric ${\bf P}_{a,i}$ with peculiar triangular restriction $1\leqslant i\leqslant a$.
This requires appropriate definition of the generalized cut-and-join operators
on this extended set of time-variables --
while the corresponding extension of Miwa transformation and
the simple formulas like above for matrix variables are not yet available.

The Yangian $Y(\hat{ \mathfrak{gl}}_1)$ is the simplest algebra in the Yangian and DIM family \cite{Tsymbaliuk:2014fvq, Maulik:2012wi,Prochazka:2015deb, Galakhov:2021xum, FJMM, NW, Mironov:2016yue,Ghoneim:2020sqi, Awata:2018svb} possessing a representation on $3d$ Young diagrams, so called MacMahon representation. Our strategy is to identify generalized cut-and-join operators with the commuting generators $\psi_n$ of the affine Yangian $Y(\hat{ \mathfrak{gl}}_1)$ \cite{Tsymbaliuk:2014fvq, Maulik:2012wi, Prochazka:2015deb}.  The purpose is to build a representation of the Yangian $Y(\hat{ \mathfrak{gl}}_1)$ as explicit differential operators in {\it triangular} variables ${\bf P}_{a \geqslant i}$, hence we call this representation {\it triangular}. Then 3-Schur polynomials are computed as common eigenfunctions of operators $\psi_n$ in explicit form.

\subsection{Motivation}

The role of Schur functions $S_R$ in modern theory is diverse and constantly increasing. They appear in the following incomplete set of areas:
\begin{itemize}
    \item $S_R$ provide explicit description of the Schur-Weyl duality between symmetric $s_N$ and linear $SL_N$ groups;
    \item $S_R$ are characters of irreducible representations;
    \item $S_R$ are central elements in the description of KP/Toda integrability --
which is the universal feature of non-perturbative partition functions
\cite{UFN3,GMMMO,GKMMM,KMMMZ, Kazakov:1998ji, MIntMM, LMMMPWZ,MMMPZ, MMMPWZ};
    \item $S_R$ are used in various character decompositions \cite{Munitary,Mironov:2011aa, Mironov:2011ym,Itoyama:2012qt};
    \item $S_R$ and their relatives are eigenfunctions of commuting set of Hamiltonians of quantum integrable many-body systems \cite{Awata:1994xd,MMintsystWLZZ, awata2000hidden, MMZellSym, Ker1Mironov:2019};
    \item $S_R$ provide the {\it superintegrable} basis \cite{MMsi,MMsidouble, MMsiNekr, Mironov:2020tjf,MM1MM, MMZ, MMMZ, Mishnyakov:2022bkg, Wang:2022fxr, Bawane:2022cpd, Cassia:2020uxy} in matrix models.
\end{itemize}

In different applications they show up either as symmetric functions $S_R[x]$
of $N$ variables $x$, which can be also treated as eigenvalues of the $N\times N$ matrices $X_{i,j}$, or as the functions of infinitely many time-variables $p_a$, which form an expansion basis for KP/Toda $\tau$-functions --
which can then be described as bilinear Plucker relations for the expansion coefficients. These two show-ups are related by Miwa transform:
\begin{equation}
    p_a=\tr X^a = \sum_{i=1}^N x_i^a
\end{equation}
which is a restriction of infinite set of time-variables to the $N$-dimensional Miwa locus -- what emphasises the more fundamental/universal role of the time-version. There are two main determinant formulas for expressing the Schur polynomials as symmetric functions:
\begin{equation}
    S_{R}[x] = \frac{\det_{1 \leqslant i,j \leqslant N} \left( x_{j}^{R_i + N - i} \right)}{\det_{1 \leqslant i,j \leqslant N} \left( x_{j}^{N - i} \right)}
\end{equation}
or in terms of time-variables:
\begin{equation}
    S_{R}[p] = \det_{1 \leqslant i,j \leqslant l(R)} \left( S_{[R_i + j - i]} \right), \hspace{15mm} \sum_{k=0}^{\infty} S_{[k]} \cdot z^{k} = \exp \left\{ \sum_{k=1}^{\infty} \frac{p_k \cdot z^{k}}{k}\right\}
\end{equation}
where one-row Schur polynomials $S_{[k]}$ are computed from the generating function.

Schur functions form a complete and orthogonal set, with orthogonality condition known as Cauchy identity \cite{Morozov:2018lsn}:
\begin{equation}
    \sum_{R} S_{R}[p] \cdot S_{R}[p^{\prime}] = \exp \left\{ \sum_{k=1}^{\infty} \frac{p_k \cdot p_{k}^{\prime}}{k}\right\}
\end{equation}
Here the sum in the l.h.s runs over all Young diagrams and $p_k$, $p^{\prime}_k$ are two idependent sets of time-variables. The orthogonality property is important for our presentation and we use it in more general form:
\begin{equation}
\label{Cauchy id}
    \sum_{R} \frac{{}S_{R}[p] \cdot S_{R}[p^{\prime}]}{|| S_{R} ||^2} = \exp \left\{ \sum_{k=1}^{\infty} \frac{p_k \cdot p_{k}^{\prime}}{|| p_k ||^2}\right\}
\end{equation}
where the normalization of Schur polynomials and time-variables are different from the classical choice.

As soon as Schur functions $S_{R}[p]$ have two kinds of variables -- Young diagram and time-varibles, this allows them to be the transformation matrices in Schur-Weyl duality, since $R$ are natural for the action of symmetric group, while linear algebras act as differential operators in times ({\it a la} free-field representations of \cite{GMMOS,Leites1,DiffOper1Morozov:2021hwr,DiffOper2Morozov:2022ocp}).

As symmetric functions and even those of times Schur functions have obvious
generalization: to $q$-deformed algebras (finite differences instead of differentials),
and further to $q,t$ and elliptic cases, related to DIM algebras \cite{Mironov:2016yue} and AGT relations
\cite{MMZagt} -- the corresponding functions are known as Jack, Hall-Littlewood \cite{Kerov},
Macdonald \cite{Macdonald, Zenkevich:2017tnb}, Shiraishi \cite{Awata:2020xfq} and elliptic Shirasishi \cite{Awata:2020yxf} functions.

Much less understood is the generalization in the other direction --
that of the Young diagrams $R$.
One of intriguing possibilities is to raise integer partitions to plane partitions, from $2d$ to $3d$ Young diagrams.
Since the number of these diagrams is counted by MacMahon formula
\be
\sum_{n} \left[\#_{\text{ plane (3d) partitions of size}\ n}\right] \cdot q^n = \prod_{k=1}^\infty \frac{1}{(1-q^k)^k} = 1 + q + 3 q^2 + 6 q^3 + 13 q^4 + 24 q^5 + 48 q^6  + \ldots
\ee
which substitutes the Dedekind function
\be
\sum_{n} \left[\#_{\text{ integer (2d) partitions of size}\ n} \right] \cdot q^n = \prod_{k=1}^\infty \frac{1}{1-q^k} = 1 + q + 2 q^2 + 3 q^3 + 5 q^4 + 7 q^5 + 11 q^6 \ldots
\ee
for integer partitions,
it is clear that the set of times $p_k$ should be changed for a triangular set ${\bf P}_{k,i}$ with $1\leqslant i\leqslant k$. In other words, each factor $\frac{1}{1-q^k}$ corresponds to a time variable of order $k$. In the $3d$ case of MacMahon formula there are $k$ such identical factors that should be identified with $k$ time-variables of equal order $k$. In our notation $3d$ time-variables are ${\bf P}_{k,i}$, where $k$ is the order and $i$ ($1 \leqslant i \leqslant k$) is a label.
\begin{table}[h!]
\ytableausetup{boxsize = 0.7em}
\doublespacing
    \centering
    \begin{tabular}{|c|c|c|}
    \hline
        1 & ${\bf P}_{1,1}$ &  \begin{ytableau} \  \\ \end{ytableau} \\
        \hline
        2 & ${\bf P}_{2,1}, \ {\bf P}_{2,2}, \ ({\bf P}_{1,1})^2$ & 
        \begin{ytableau} \ & \ \\ \end{ytableau}, \begin{ytableau} 
        \ \\
        \ \\
        \end{ytableau}, \begin{ytableau}
         \yt{2} \\
        \end{ytableau}\\
        \hline
        3 & ${\bf P}_{3,1}, \ {\bf P}_{3,2}, \ {\bf P}_{3,3}, \ {\bf P}_{2,1} {\bf P}_{1,1},  \ {\bf P}_{2,2} {\bf P}_{1,1}, \ ({\bf P}_{1,1})^3$ &
        \begin{ytableau} \ & \ & \ \\ \end{ytableau}, \begin{ytableau} \ & \ \\
        \ \\
        \end{ytableau},
        \begin{ytableau}
        \ \\
        \ \\ 
        \ \\
        \end{ytableau}, 
        \begin{ytableau}
        \yt{2} & \
        \end{ytableau}, 
        \begin{ytableau}
        \yt{2} \\
        \
        \end{ytableau}, 
        \begin{ytableau}
        \yt{3} 
        \end{ytableau}\\
        \hline
        4 & ${\bf P}_{4,1}, \ {\bf P}_{4,2}, \ {\bf P}_{4,3}, \ {\bf P}_{4,4}, \ ({\bf P}_{2,1})^2, \ {\bf P}_{2,1}{\bf P}_{2,2}, \ ({\bf P}_{2,2})^2,$ & \begin{ytableau}
            \ & \ & \ & \ 
        \end{ytableau}, \begin{ytableau}
            \ & \ & \ \\
            \
        \end{ytableau},  \begin{ytableau}
            \ & \ \\
            \ & \ 
        \end{ytableau},  \begin{ytableau}
            \ & \ \\
            \ \\
            \
        \end{ytableau}, \begin{ytableau}
            \ \\
            \ \\
            \ \\
            \
        \end{ytableau}, \\
        & $ {\bf P}_{3,1}{\bf P}_{1,1}, \ {\bf P}_{3,2}{\bf P}_{1,1}, \ {\bf P}_{3,3}{\bf P}_{1,1}, \ {\bf P}_{2,1} ({\bf P}_{1,1})^2, \ {\bf P}_{2,2} ({\bf P}_{1,1})^2, \ ({\bf P}_{1,1})^4$ & \begin{ytableau}
            \yt{2} & \ & \ \\
        \end{ytableau}, 
        \begin{ytableau}
            \yt{2} \\
            \ \\
            \ \\
        \end{ytableau}, 
        \begin{ytableau}
            \yt{2} & \yt{2} 
        \end{ytableau},
        \begin{ytableau}
            \yt{2} \\
            \yt{2}
        \end{ytableau},
        \begin{ytableau}
            \yt{3} & \ 
        \end{ytableau},
        \begin{ytableau}
            \yt{3} \\
            \
        \end{ytableau},
        \begin{ytableau}
            \yt{2} & \ \\
            \ 
        \end{ytableau},
        \begin{ytableau}
            \yt{4}
        \end{ytableau} \\
        \hline
    \end{tabular}
    \label{tab:3d Young diagrams}
    \caption{\scriptsize{The number of $3d$ Young diagrams with $n$ boxes is equal to the number of homogeneous monomials of times ${\bf P}_{a,i}$, where the grading of variable ${\bf P}_{a,i}$ is equal to $a$. Our notation of the $3d$ Young diagrams is the following: the numbers in the boxes are the heights and the empty box means unit height. Note that in $3d$ case there is no natural correspondence between monomials and the diagrams.}}
\end{table}

\begin{table}[h!]
\ytableausetup{boxsize = 0.7em}
\doublespacing
    \centering
    \begin{tabular}{|c|c|c|}
    \hline
        1 & $p_1$ &  \begin{ytableau} \  \\ \end{ytableau} \\
        \hline
        2 & $p_2, (p_1)^2$ &  \begin{ytableau} \ & \ \\ \end{ytableau}, \begin{ytableau} \ \\
        \ \\\end{ytableau} \\
        \hline
        3 & $p_3, p_2 p_1, (p_1)^3$ & \begin{ytableau} \ & \ & \ \\ \end{ytableau}, \begin{ytableau} \ & \ \\
        \ \\
        \end{ytableau},
        \begin{ytableau}
        \ \\
        \ \\ 
        \ \\
        \end{ytableau} \\
        \hline
        4 & $p_4, p_3 p_1, (p_2)^2, p_2 (p_1)^2, (p_1)^4$ & 
        \begin{ytableau} 
        \ & \ & \ & \ \\ 
        \end{ytableau}, 
        \begin{ytableau} 
        \ & \ & \ \\
        \ \\
        \end{ytableau},
        \begin{ytableau}
        \ & \ \\
        \ & \ \\
        \end{ytableau},
        \begin{ytableau}
        \ & \ \\
        \ \\
        \ \\
        \end{ytableau},
        \begin{ytableau}
        \ \\
        \ \\ 
        \ \\
        \ \\
        \end{ytableau} \\
        \hline
    \end{tabular}
    \caption{\scriptsize{The number of $2d$ Young diagrams with $n$ boxes is equal to the number of homogeneous monomials of times $p_{a}$, where the grading of variable $p_a$ is equal to $a$.}}
    \label{tab:2d Young diagrams}
\end{table}
\subsection{Plan}
All our previous attempts \cite{Morozov:2018fga,Morozov:2018fjb,MThunt} to construct 3-Schur polynomials are failed. In this paper we use the Yangian $Y(\hat{ \mathfrak{gl}_1})$ to attack this problem for the following reason: the Yangian is the proper 2-parameter deformation of the $W_{1 + \infty}$ algebra, that possess so called MacMahon representation, which vectors are enumerated by the $3d$ Young diagrams \cite{Prochazka:2015deb,Tsymbaliuk:2014fvq}. Our strategy is described by three steps:
\begin{enumerate}
    \item {\bf Find triangular representation of the Yangian} $Y(\hat{ \mathfrak{gl}_1})$ in terms of differential operators in time-variables ${\bf P}_{a \geqslant i}$.
    \item Compute 3-Schur polynomials ${\bf S}_{\pi}$ as the mutual eigenfunctions of commuting subalgebra of generalized cut-and-join operators $\psi_k$.
    \item Check the basic properties of the obtained 3-Schur polynomials: $2d$ limit and orthogonality properties via Cauchy identity.
\end{enumerate}
The paper is organized as follows. In Sect. \ref{Yang} we review basic facts about the Yangian and discuss the distinguished role of $\psi_3$ operator. In Sect. \ref{2d representation} we discuss $2d$ representation of the Yangian that corresponds to the Jack and 2-Schur polynomials. In the main Sect. \ref{3d representation} and \ref{Examples} we construct triangular Yangain representation and compute 3-Schur polynomials level by level. Finally, in Sect. \ref{conc} we discuss the results.

\bigskip

\section{Yangian of $\hat{\mathfrak{gl}}_1$}
\label{Yang}
We define the Yangian following the presentations of \cite{Tsymbaliuk:2014fvq, Prochazka:2015deb}. The algebra possesses three types of generators $e_i, f_i, \psi_i$ where $i = 0,1,2,\ldots$ that obey following set of relations: 
\begin{align}
\phantom. [\psi_j,\psi_k]&=0 \\
\phantom. [e_i,f_j] &= \psi_{i+j}
\label{linear relations}
\end{align}
The next set of relations contains an anti-commutator term:
\begin{align}
\begin{aligned}
\phantom. [e_{j+3},e_k] - 3[e_{j+2},e_{k+1}] + 3[e_{j+1},e_{k+2}] - [e_j,e_{k+3}] &= O(h)
= \sigma_2\Big([e_j,e_{k+1}]-[e_{j+1},e_k]\Big)+\sigma_3\{e_j,e_k\}   \\
\phantom. [f_{j+3},f_k] - 3[f_{j+2},f_{k+1}] + 3[f_{j+1},f_{k+2}] - [f_j,f_{k+3}] &= O(h)
= \sigma_2\Big([f_j,f_{k+1}]-[f_{j+1},f_k]\Big)-\sigma_3\{f_j,f_k\}   \\
\phantom. [\psi_{j+3},e_k] - 3[\psi_{j+2},e_{k+1}] + 3[\psi_{j+1},e_{k+2}] - [\psi_j,e_{k+3}] &= O(h)
= \sigma_2\Big([\psi_j,e_{k+1}]-[\psi_{j+1},e_k]\Big)+\sigma_3\{\psi_j,e_k\}  \\
\phantom. [\psi_{j+3},f_k] - 3[\psi_{j+2},f_{k+1}] + 3[\psi_{j+1},f_{k+2}] - [\psi_j,f_{k+3}] &= O(h)
= \sigma_2\Big([\psi_j,f_{k+1}]-[\psi_{j+1},f_k]\Big)-\sigma_3\{\psi_j,f_k\}
\end{aligned}
\label{quadratic relations}
\end{align}
The third set of relations are Serre-like relations:
\begin{align}
\begin{aligned}
\label{Serre rel}
    {\rm Sym}_{j_1,j_2,j_3} \left[e_{j_1},[e_{j_2},e_{j_3+1}]\right] &= 0  \\
    {\rm Sym}_{j_1,j_2,j_3} \left[f_{j_1},[f_{j_2},f_{j_3+1}]\right] &= 0
\end{aligned}
\end{align}

The last set of relations fixes the commutation relation with $\psi_{0},\psi_{1},\psi_{2}$:
\begin{align}
    \begin{aligned}
        \relax [\psi_0, e_i ] &= 0 &\hspace{15mm} [\psi_1, e_i] &= 0 &\hspace{15mm} [\psi_2, e_i] &= 2 e_i \\
        [\psi_0, f_i ] &= 0 &\hspace{15mm} [\psi_1, f_i] &= 0 &\hspace{15mm} [\psi_2, f_i] &= -2 f_i \\
    \end{aligned}
    \label{boundary conditions}
\end{align}
These relations means that $\psi_0, \psi_1$ could be thought as numbers and $\psi_2$ is a diagonal grading operator. 

The algebra has three parameters $h_1, h_2, h_3$ that are complex numbers obeying $h_1+h_2+h_3=0$
Deformation parameters $\sigma_2, \sigma_3$ that enter the algebra relations \eqref{quadratic relations} are symmetric functions of parameters $h_1,h_2,h_3$:
\begin{align}
    \begin{aligned}
        \sigma_2 &:= -\frac{1}{2}(h_1^2+h_2^2+h_3^2) = h_1h_2+h_2h_3+h_3h_1 \\
        \sigma_3 &:=h_1h_2h_3
    \end{aligned}
\end{align}  

Note that effectively there are only two independent parameters $h_i$, therefore one should compare formulas keeping this fact in mind.

The generators could be summed into generating functions:
\begin{align}
    e(u) &:= \sum_{i=0}^{\infty} \frac{e_i}{u^{i+1}} \hspace{25mm}
    \psi(u) :=1 + \sigma_3 \sum_{i=0}^{\infty} \frac{\psi_i}{u^{i+1}} \hspace{25mm}
    f(u) := \sum_{i=0}^{\infty} \frac{f_i}{u^{i+1}} 
    \label{generating functions}
\end{align}

\bigskip
The Yangian has special set of operators
\begin{align}
    \boxed{\boxed{e_0, \psi_3, f_0}}
    \label{special set}
\end{align}
with a remarkable property that we extensively exploit in this paper: the other Yangian generators can be obtained from them. In particular, using third and fourth relation of \eqref{quadratic relations} for $j=0$ and \eqref{boundary conditions} one obtains:
\begin{align}
    \begin{aligned}
        e_{k+1} = \ &\frac{1}{6} [ \psi_3, e_{k} ] - \frac{1}{3} \psi_0 \sigma_3 \, e_{k} \\
        f_{k+1} = -&\frac{1}{6} [ \psi_3, f_{k} ] - \frac{1}{3} \psi_0 \sigma_3 \, f_{k} 
    \end{aligned}
    \label{higher e and f from special set}
\end{align}
These equations allow one to express any generator $e_i,f_i$ through small set of generators \eqref{special set} and complex number $\psi_0$. Then higher generators $\psi_k$ could be obtained from $e_i,f_i$ via relations \eqref{linear relations}. Based on this remarkable property of the Yangian, one can see that it is enough to construct self-consistent representation of three special operators \eqref{special set} to construct the representation of the full Yangian. 

\bigskip

\section{$2d$ representation of Yangian in terms of ordinary times $p_a$}
\label{2d representation}
The representation of Yangian is in terms of the ordinary time-variables $p_a$ is well known in the literature \cite{Prochazka:2015deb} and corresponds to the special locus of Yangian parameters $h_1 = -h, h_2 =  1/h, h_3 = h-1/h$. In the special case $h=1$ the Yangian is identified with $W_{1+\infty}$-algebra. The whole representation is recovered from the cut-and-join operator:
\begin{align}
    \boxed{
    \psi_3 = 3\sum_{a,b = 1}^{\infty} \left(ab \cdot p_{a+b}\frac{\p^2}{\p p_a \p p_b} + (a+b) \cdot p_ap_b\frac{\p}{\p p_{a+b}}\right) + \sigma_3 \sum_{a=1}^{\infty} a(3a-1) p_a \frac{\p}{\p p_a}} 
    \label{2d cut and join}
\end{align}
We provide the explicit form of several $e_k$ and $f_k$ operators:
\begin{equation}
    \boxed{e_0 = p_1} 
\end{equation}
\begin{equation}
    e_1 = \frac{1}{6}[\psi_3,e_0] - \frac{1}{3}\sigma_3 e_0 = \sum_{a=1}^{\infty} a \cdot p_{a+1}\frac{\p}{\p p_a} 
\end{equation}
\begin{equation}
    e_2 = \frac{1}{6}[\psi_3,e_1] - \frac{1}{3}\sigma_3 e_1 =
\sum_{a,b = 1}^{\infty} \left(ab \cdot p_{a+b+1} \frac{\p^2}{\p p_a\p p_b}+(a+b-1) \cdot p_ap_b\frac{\p}{\p p_{a+b-1}}\right)
+\sigma_3 \sum_{a=1}^{\infty} a^2 \cdot p_{a+1}\frac{\p}{\p p_a}
\end{equation}
\begin{align}
\begin{aligned}
e_3 &= \frac{1}{6}[\psi_3,e_2] - \frac{1}{3}\sigma_3 e_2 =
\sum_{a,b,c=1}^{\infty}\left(abc \cdot p_{a+b+c+1} \frac{\p^3}{\p p_a\p p_b \p p_c}+(a+b+c-1) \cdot p_ap_bp_c\frac{\p}{\p p_{a+b+c-1}}\right)
+ \\
&+\frac{3}{2} \sum_{a,b,c,d=1}^{\infty} \delta_{a+b,c+d+1} \cdot cd \cdot p_ap_b\frac{\p^2}{\p p_c\p p_d} + \frac{1}{2} \sum_{a=1}^{\infty} (a+1+2a\sigma_3^2)a^2 \cdot p_{a+1}\frac{\p}{\p p_a}\\
&+\frac{\sigma_3 }{2}\sum_{a,b=1}^{\infty} \left(3ab(a+b) \cdot p_{a+b+1} \frac{\p^2}{\p p_a\p p_b}
+\big(3(a+b)-4\big)(a+b-1) \cdot p_ap_b\frac{\p}{\p p_{a+b-1}}\right)
\end{aligned}
\end{align}

\begin{equation}
    \boxed{ f_0 = -\frac{\p}{\p p_1}}\\
\end{equation}
\begin{equation}
    f_1 = -\frac{1}{6}[\psi_3,f_0] - \frac{1}{3}\sigma_3 f_0 = - \sum_{a=1}^{\infty} (a+1) \cdot p_{a}\frac{\p}{\p p_{a+1}} \\
\end{equation}
\begin{equation}
    f_2 = -\frac{1}{6}[\psi_3,f_1] - \frac{1}{3}\sigma_3 f_1 =
-\sum_{a,b=1}^{\infty}\left(ab \cdot p_{a+b-1} \frac{\p^2}{\p p_a\p p_b}+(a+b+1) \cdot p_ap_b\frac{\p}{\p p_{a+b+1}}\right)
-\sigma_3 \sum_{a=1}^{\infty }(a+1)a \cdot p_{a}\frac{\p}{\p p_{a+1}}
\end{equation}
\begin{align}
    \begin{aligned}
    f_3 &=  -\frac{1}{6}[\psi_3,f_2] - \frac{1}{3}\sigma_3 f_2 =
- \sum_{a,b,c=1}^{\infty}\left(abc \cdot p_{a+b+c-1} \frac{\p^3}{\p p_a\p p_b \p p_c}+(a+b+c+1) \cdot p_ap_bp_c\frac{\p}{\p p_{a+b+c+1}}\right) -\\
&- \frac{3}{2} \sum_{a,b,c,d=1}^{\infty} \delta_{a+b,c+d-1} \cdot cd \cdot p_ap_b\frac{\p^2}{\p p_c\p p_d} - \frac{1}{2} \sum_{a=1}^{\infty} (a+1+2a\sigma_3^2)(a+1)a \cdot p_{a}\frac{\p}{\p p_{a+1}}\\
&-\frac{\sigma_3 }{2}\sum_{a,b=1}^{\infty} \left(\big(3(a+b)-4\big)ab \cdot p_{a+b-1} \frac{\p^2}{\p p_a\p p_b}
+3(a+b)(a+b+1) \cdot p_ap_b\frac{\p}{\p p_{a+b+1}}\right)
    \end{aligned}
\end{align}

Then operators $\psi_k$ can be computed:
\begin{equation}
    \psi_0 = [e_0,f_0]=1
\end{equation}
\begin{equation}
    \psi_1 = [e_1,f_0] = [e_0,f_1] = 0
\end{equation}
\begin{equation}
    \psi_2 = [e_1,f_1] = [e_2,f_0] = [e_0,f_2] = 2\sum_{a=1}^{\infty} a \cdot p_a \frac{\p}{\p p_a} \\
\end{equation}

The next relations are self-consistency checks for the representation:
\begin{equation}
    \psi_3 = [e_3,f_0] = [e_2,f_1] = [e_1,f_2] = [e_0,f_3]
\end{equation}

In fact only the three expressions for $e_0,f_0, \psi_3$ operators (in the boxes) are needed -- all the rest are their corollaries
and we list just a few, because they are recognizable in many other areas
(see \cite{MMMPZ} for the latest example).
Since two expressions for $e_0$ and $f_0$ are just trivial, the truly needed is the single expression for the cut-and-join operator $\psi_3$ -- all the rest follows.
However, the possible expression for $\psi_3$ is strongly constrained by the highly non-trivial
Yangian relations :  commutativity of all (infinitely many) $\psi_k$ \eqref{linear relations}, quadratic Yangian relations \eqref{quadratic relations} and Serre relations \eqref{Serre rel}
for $e_k$ and $f_k$.

It is far from obvious that above $\psi_3$ satisfies all this infinite set of relations, but it does. 
The only restriction is that $\sigma_2$ is not arbitrary, but is expressed through $\sigma_3$:
\be
\sigma_3^2 + \sigma_2 +1 = 0
\label{2dconstr}
\ee
In general this 1-parametric family of representations  describes Jack polynomials for $h_1 = -h, h_2 =  1/h, h_3 = h-1/h$,
in the case of $\sigma_3=0, h = 1$ they are reduced to $2$-Schurs. We provide examples of Jack polynomials, that can be directly computed as eigenfunctions of \eqref{2d cut and join}:
\begin{equation}
    \psi_3 \, J_{R} = \lambda_R \, J_R \hspace{15mm} \lambda_R = \sum_{\Box \in R} \left( \frac{6}{h} \, x_2(\Box) - 6h x_1(\Box) + 2 \sigma_3 \right)
\end{equation}
\begin{align}
    \begin{aligned}
        \lambda_{\, \begin{ytableau}
            \
        \end{ytableau}} &= 2 \left( h - \frac{1}{h} \right) &\hspace{15mm} J_{\, \begin{ytableau}
            \
        \end{ytableau}} &= p_1 \\
        \lambda_{\, \begin{ytableau}
            \ \\ 
            \
        \end{ytableau}} &= -6h + 4 \left( h - \frac{1}{h} \right) &\hspace{15mm} J_{\, \begin{ytableau}
            \ \\
            \
        \end{ytableau}} &= p_1^2 - h p_2 \\
        \lambda_{\, \begin{ytableau}
            \ & \
        \end{ytableau}} &= \frac{6}{h}+ 4 \left( h - \frac{1}{h} \right) &\hspace{15mm} J_{\, \begin{ytableau}
            \ & \
        \end{ytableau}} &= p_1^2 + \frac{1}{h} p_2 \\
    \end{aligned}
\end{align}

 What we need is generalization of the formula for $\psi_3$
to the $3d$ representation in terms of ${\bf P}_{a \geqslant i}$,
where the restriction (\ref{2dconstr}) will be lifted and $\sigma_2,\sigma_3$ become
two independent deformation parameters and we get the full-fledged 2-parametric family
of $3$-Schur functions.

\bigskip

\section{Triangular $3d$ representation of Yangian. Outline of the algorithm}
\label{3d representation}
Our algorithm is based on the peculiar feature of the Yangian: its representation can be completely fixed by self-consistent set op operators \eqref{special set}. Recall, the structure of the cut-and-join operator for $2d$ case:
\begin{align}
    \begin{aligned}
        \psi^{2d}_3 &= \sum_{a} U^a \cdot p_a \frac{\p }{\p p_a} + \sum_{a,b} X^{a,b} \cdot p_{a+b} \frac{\p^2}{\p p_a \p p_b} + Y^{a,b} \cdot p_a p_b \frac{\p }{\p p_{a+b}}
        \label{2d ansatz}
    \end{aligned}
\end{align}
We do not specify here the explicit form of functions $U^a, X^{a,b}, Y^{a,b}$ to draw the attention to the structure of these operators. Namely, the sum of indices that are "cut" with derivatives is always equal to the sum of indices that are "inserted" by multiplication in the cut-and-join operator $\psi^{2d}_3$. This property is important for out presentation and we generalize it in the following way to the $3d$ case.

Our starting point for the triangular representation is the ansatz for cut-and-join operator, that is the direct generalization of \eqref{2d ansatz} according to the rule $p_{a} \to {\bf P}_{a,i}$:
\begin{equation}
\boxed{\boxed{
    \psi_3^{\text{triang}} = \sum_{a=1}^{\infty} \sum_{i,j=1}^a \, U^a_{i,j} \cdot {\bf P}_{a,i}\frac{\p}{\p {\bf P}_{a,j}} +
\sum_{a,b=1}^{\infty} \sum_{i=1}^a \sum_{j=1}^b \sum_{k=1}^{a+b} \left(X^{a,b}_{i,j|k} \cdot {\bf P}_{a+b,k} \frac{\p^2}{\p {\bf P}_{a,i}\p {\bf P}_{b,j}}
+ Y^{a,b}_{i,j|k} \cdot {\bf P}_{a,i}{\bf P}_{b,j}\frac{\p}{\p {\bf P}_{a+b,k}}\right)
}}
\label{3d cut and join}
\end{equation}

Further in the paper we omit the superscript {\it triangular} for the Yangain operators. The functions $U^{a}_{i,j}$, $X^{a,b}_{i,j|k}$, $Y^{a,b}_{i,j|k}$ receive additional indices corresponding to the second indices of times ${\bf P}_{a,i}$ and obey obvious symmetries:
\begin{align}
    \begin{aligned}
        X^{a,b}_{i,j|k} = X^{b,a}_{j,i|k} \hspace{25mm}
        Y^{a,b}_{i,j|k} = Y^{b,a}_{j,i|k}\\
    \end{aligned}
\end{align}

The operators $e_0,f_0$ are generalized according to the same rule:
\begin{align}
\boxed{\boxed{
    \begin{aligned}
        e_0 &= \psi_0 \, {\bf P}_{1,1} \\
        f_0 &= -\frac{\p }{\p {\bf P}_{1,1}}
    \end{aligned}
    }}
\end{align}

Here $\psi_0$ is a complex number that characterize the representation.
Recall that all the Yangian generators can be obtained from the special set \eqref{special set} via relations \eqref{higher e and f from special set}. For our ansatz the Yangain generators are functions of coefficients $U^{a}_{i,j}, X^{a,b}_{i,j|k}, Y^{a,b}_{i,j|k}$. The Yangian relations are not satisfied for arbitrary values of $U^{a}_{i,j}, X^{a,b}_{i,j|k}, Y^{a,b}_{i,j|k}$ but impose special conditions on them. Further we explain how to find consistent solution, however we should emphasize one important point. 

We stress that the ansatz \eqref{3d cut and join} is a {\it nontrivial} conjecture since it is not guaranteed that higher terms like $ {\bf P}_{a,i} {\bf P}_{b,j} {\bf P}_{c,k} \frac{\p }{\p {\bf P}_{a+b+c,l}}$ will not be required to solve the Yangian relations from Section \ref{Yang}.

We solve the Yangian relations level by level and then find the 3-Schur polynomials on the corresponding level as the eigenfunctions of the $\psi_3$ operator. For each level $n$ we follow the following scheme:
\begin{enumerate}
    \item First, we fix the basis in the subspace of times ${\bf P}_{n,i}$ for $i = 1, \ldots, n$. All these times have the same grading and therefore can be mixed together via linear transformations. We use this freedom to impose the following conditions:
    \begin{align}
        \begin{aligned}
            Y_{1,j|k}^{1,b} \sim \delta_{j,k} \hspace{25mm}
            X_{1,j|k}^{1,b} \sim \delta_{j,k}
        \end{aligned}
    \end{align}
    
    In other words the coefficients $Y_{1,j|k}^{1,b}, X_{i,j|k}^{1,b}$ vanish unless $k=j$. The values of $Y_{1,j|j}^{1,b}, X_{i,j|j}^{1,b}$ is determined from the next steps. These conditions are motivated by the $2d$ case, where all the coefficients $X_{i,j|k}^{a,b}, Y_{i,j|k}^{a,b}$ obey the rule $k = i+j-1$ since the second index of time-variables always equal to one, i.e. $i = j = k = 1$. Therefore we choose the condition $k = i+j-1$ as the defining property of our algorithm. It does not mean that there are not better choices. 
    \item Next we impose the Yangain relation:
    \begin{equation}
        [ \psi_2, \psi_3 ] = 0
    \end{equation}
    That are equivalent to relations:
    \begin{equation}
        X_{1,j|j}^{1,b} \, Y_{1,j|j}^{1,b} = \frac{3(b-j+1)}{\psi_0} \cdot 3(b+j)
    \end{equation}
    We make natural choice:
    \begin{align}
        \begin{aligned}
            X_{1,j|j}^{1,b} = \frac{3(b-j+1)}{\psi_0} \hspace{25mm}
            Y_{1,j|j}^{1,b} = 3(b+j)
        \end{aligned}
        \label{times normalization}
    \end{align}
    This choice corresponds to particular choice of scalar products of time-variables in Cauchy identity \eqref{Cauchy id}.
    \item Next we impose higher relations to fix the remaining coefficients  $U^{a}_{i,j}, X^{a,b}_{i,j|k}, Y^{a,b}_{i,j|k}$:
    \begin{align}
        \begin{aligned}
            \psi_3 -[e_2, f_1] &=0  \\
            \text{YRel}_{j,k}:=[e_{j+3},e_k] - 3[e_{j+2},e_{k+1}] + 3[e_{j+1},e_{k+2}] - [e_j,e_{k+3}] -\sigma_2\Big([e_j,e_{k+1}]-[e_{j+1},e_k]\Big)-\sigma_3\{e_j,e_k\} &=0  \\
            {\rm Sym}_{j_1,j_2,j_3} \left[e_{j_1},[e_{j_2},e_{j_3+1}]\right] &= 0  \\
        \end{aligned}
        \label{Yangian relations}
    \end{align}
    \item When all the coefficients $U^{a}_{i,j}, X^{a,b}_{i,j|k}, Y^{a,b}_{i,j|k}$ are fixed on the level $n$ one can compute the 3-Schur polynomials ${\bf S}_{\pi}$ as it's eigenvectors:
    \begin{equation}
        \psi_3 \, {\bf S}_{\pi} = \lambda_{\pi} {\bf S_{\pi}}
    \end{equation}
    The eigenvectors are distinguished by the eigenvalues $\lambda_{\pi}$ that obey conjectural formula:
    \begin{equation}
         \lambda_{\pi} =  \sum_{\Box \in \pi} 6 h_{\Box} + 2 \sigma_3 \psi_0
         \label{eigenvalues}
    \end{equation}
    Here the content-function $h_{\Box}$ and it is defined as follows:
\begin{equation}
    h_{\Box} := h_1 \, x_1(\Box) + h_2 \, x_2(\Box) + h_3 \, x_3(\Box)
\end{equation}
where $x_i(\Box)$ are coordinates of the box, the coordinate of the box in the single box Young diagram is set to be (0,0,0).
\end{enumerate}
\bigskip

Now we make few comments on the choice of complex parameter $\psi_0$. One can perform calculations for arbitrary value of $\psi_0$, however the expressions quickly become too lengthy. Different choices such as  $\psi_0 = 1$  are presented in the literature\cite{Prochazka:2015deb, Wang1}. We argue that the following choice
\begin{equation}
\boxed{
    \psi_0 = -\frac{1}{\sigma_2}
    }
\end{equation}
has two main advantages: the formulas are significantly simplified and the homogeneity is restored. By the restoration we mean the appearance of an additional grading:
\begin{align}
\begin{aligned}
     \left[h_i\right] &= 1  &\hspace{15mm} \left[\sigma_2\right] &= 2 &\hspace{15mm} \left[\sigma_3\right] &= 3 \\
     \left[{\bf P}_{a,i}\right] &= 1 &\hspace{15mm} \left[\frac{\p }{\p {\bf P}_{a,i}}\right] &= -1 & &\\
     \left[e_i\right] &= i-1 & \hspace{15mm} \left[\psi_i\right] &= i-2 &\hspace{15mm} \left[f_i\right] &= i-1\\
\end{aligned}
\label{grading}
\end{align}
For the first few levels we provide both formulas: for arbitrary value $\psi_0$ and $\psi_0 = -1/\sigma_2$, while for higher levels we provide only simplified version for $\psi_0 = -1/\sigma_2$. 
\bigskip

\section{Examples at low levels}
\label{Examples}
\setcounter{subsection}{-1} 
\subsection{0 level}
On this level there is only one vector corresponding to empty partition and we choose it to be one:
\begin{equation}
    {\bf S}_{\varnothing} = 1
\end{equation}
According to rules \eqref{higher e and f from special set} one can compute $e_1, f_1$ and then compute the complex number $\psi_1$:
\begin{equation}
    \psi_1 = U_{1,1}^{1} - 2 \sigma_3 \psi_0
\end{equation}
Alternatively, the value of $\psi_1$ can be fixed by the following considerations. One computes $e_2$ and substitute it to the following Yangian relation acting on one:
\begin{equation}
    \psi_3 \cdot 1 = [e_2, f_1] \cdot 1
\end{equation}
The l.h.s vanishes from \eqref{3d cut and join} while the r.h.s can be computed:
\begin{equation}
    0 = \frac{\psi_0}{36} \left( U_{1,1}^{1} - 2 \sigma_3 \psi_0 \right)^2
\end{equation}
We conclude:
\begin{align}
    U_{1,1}^{1} = 2 \sigma_3 \psi_0 \hspace{25mm}
    \psi_1 = 0
\end{align}

\bigskip

\subsection{1 level}
On the first level there is one diagram $\begin{ytableau}
    \
\end{ytableau}$ and one basis monomial ${\bf P}_{1,1}$. Our normalization looks as follows:
\begin{equation}
    {\bf S}_{\,\begin{ytableau}
        \
    \end{ytableau}} = {\bf P}_{1,1}
\end{equation}
The $\psi_3$ operator \eqref{3d cut and join} acts on the first level according to the following table:
\begin{equation}
     \setstretch{1.5}
        \begin{array}{c||c}
        \doublespacing
             \psi_{3} & \psi_{3}\cdot  {\bf P}_{1,1} \\
             \hline
             \hline
            {\bf P}_{1,1} & U_{1,1}^{1} = 2 \sigma_3 \psi_0
        \end{array}
\end{equation}
One can check formula \eqref{eigenvalues} for the eigenvalues:
\begin{equation}
    \lambda_{\, \begin{ytableau}
        \
    \end{ytableau}} =  2 \sigma_3 \psi_0
\end{equation}

\bigskip

\subsection{2 level}
On the second level there are three basis monomials $({\bf P}_{1,1})^2, {\bf P}_{2,1}, {\bf P}_{2,2}$ and the set of the Young diagrams contains three elements:
\begin{equation}
    \begin{ytableau}
        \ \\ \
    \end{ytableau},
    \begin{ytableau}
        \ & \
    \end{ytableau},
    \begin{ytableau}
        \yt{2}
    \end{ytableau}
\end{equation}
To completely fix the form of 3-Schur polynomials on the level two one should compute the coefficients appearing in the following table:
\begin{equation}
    \setstretch{2.5}
         \begin{array}{c||c|c|c}
             \psi_{3} & \psi_{3}\cdot{\bf P}_{1,1}^2 & \psi_{3}\cdot{\bf P}_{2,1} & \psi_{3}\cdot{\bf P}_{2,2}\\
             \hline
             \hline
             {\bf P}_{1,1}^2 &  2 U_{1,1}^{1} & Y^{1,1}_{1,1|1} & Y^{1,1}_{1,1|2} \\
             \hline
             {\bf P}_{2,1} & 2X^{1,1}_{1,1|1} & U_{1,1}^{2} & U_{1,2}^{2} \\
             \hline
             {\bf P}_{2,2} & 2X^{1,1}_{1,1|2} & U_{2,1}^{2} & U_{2,2}^{2} \\
        \end{array}
\end{equation}
that represents the action of $\psi_3$ operator on the basis monomials in the first row. The freedom of choice of basis in variables ${\bf P}_{2,1}$  and ${\bf P}_{2,2}$ allow one to impose the constrains:
\begin{align}
    \begin{aligned}
        X_{1,1|2}^{1,1} = 0 \hspace{25mm}
        Y_{1,1|2}^{1,1} = 0
    \end{aligned}
\end{align}
Next, the Yangian relation
\begin{equation}
    [\psi_2, \psi_3] = 0
\end{equation}
can not give valuable information on this low level, therefore we use higher relation acting on the basis monomial ${\bf P}_{1,1}^2$:
\begin{equation}
    (\psi_3 - [e_2,f_1])\cdot {\bf P}_{1,1}^2 = 0
\end{equation}
Solving the above relation we derive:
\begin{align}
    \begin{aligned}
         X_{1,1|1}^{1,1} = \frac{3}{\psi_0} \hspace{25mm}
         Y_{1,1|1}^{1,1} = 6 \hspace{25mm}
         U_{1,1}^{2} = 10 \sigma_3 \psi_0
    \end{aligned}
\end{align}
where we choose the normalization of time ${\bf P}_{2,1}$ according to \eqref{times normalization}. Next we use the following Yangain relation acting on $1$:
\begin{equation}
    \left([e_{3},e_0] - 3[e_{2},e_{1}] + 3[e_{1},e_{2}] - [e_0,e_{3}] -\sigma_2\Big([e_0,e_{1}]-[e_{1},e_0]\Big)-\sigma_3\{e_0,e_0\} \right) \cdot 1 =0  \\
\end{equation}
and obtain the following answer:
\begin{equation}
    U_{2,2}^2 = - 2 \sigma_3 \psi_0 \hspace{25mm} U_{2,1}^{2} U_{1,2}^2 = -\frac{36}{\psi_0} \left( 1 + \sigma_2 \psi_0 + \sigma_3^2 \psi_0^3 \right)
\end{equation}
The two coefficients $U_{1,2}^2$ and $U_{2,1}^2$ always enter into the Yangian relations in pair and one can not find their values separately. In contrast, the ratio $U_{1,2}^2 / U_{2,1}^2$ is a free parameter corresponding to the normalization of time-variable ${\bf P}_{2,2}$. We make the following symmetric choice of the normalization:
\begin{equation}
    U_{1,2}^2 = -6 \, \sqrt{ \frac{1}{\psi_0} + \sigma_2 + \sigma_3^2 \psi_0^2 } \hspace{25mm} U_{2,1}^2 = 6 \, \sqrt{ \frac{1}{\psi_0} + \sigma_2 + \sigma_3^2 \psi_0^2 }
\end{equation}
At this stage we completely fixed the form of $\psi_3$ operator on the second level:
\begin{equation}
    \setstretch{2.5}
         \begin{array}{c||c|c|c}
             \psi_{3} &  \psi_{3}\cdot{\bf P}_{1,1}^2 & \psi_{3}\cdot{\bf P}_{2,1} & \psi_{3}\cdot{\bf P}_{2,2}\\
             \hline
             \hline
             {\bf P}_{1,1}^2 &  4 \sigma_3 \psi_0 & 6 & 0 \\
             \hline
             {\bf P}_{2,1} & \frac{6}{\psi_0} & 10 \sigma_3 \psi_0 & -6 \sqrt{ \frac{1}{\psi_0} + \sigma_2 + \sigma_3^2 \psi_0^2 }\\
             \hline
             {\bf P}_{2,2} & 0 & 6 \sqrt{ \frac{1}{\psi_0} + \sigma_2 + \sigma_3^2 \psi_0^2 } & -2 \sigma_3 \psi_0 \\
        \end{array}
\end{equation}
and we are able to compute 3-Schur polynomials as the eigenvectors:
\begin{align}
    \begin{aligned}
        \lambda_{\,\begin{ytableau}
            \ \\ \
        \end{ytableau}} &= 6 h_1 + 4 \sigma_3 \psi_0, &\hspace{15mm} 
        {\bf S}_{\,\begin{ytableau}
            \ \\ \
        \end{ytableau}} &= {\bf P}_{1,1}^2 + h_1 \, {\bf P}_{2,1} + \sqrt{\frac{(1 + h_1 h_2 \psi_0)(1+ h_1 h_3 \psi_0)}{\psi_0(1 + h_2 h_3 \psi_0)}} \, {\bf P}_{2,2} \\
        \lambda_{\,\begin{ytableau}
            \ & \
        \end{ytableau}} &= 6 h_2 + 4 \sigma_3 \psi_0, &\hspace{15mm}{\bf S}_{\,\begin{ytableau}
            \ & \
        \end{ytableau}} &= {\bf P}_{1,1}^2 + h_2 \, {\bf P}_{2,1} +\sqrt{\frac{(1 + h_2 h_1 \psi_0)(1+ h_2 h_3 \psi_0)}{\psi_0(1 + h_1 h_3 \psi_0)}} \, {\bf P}_{2,2} \\
        \lambda_{\,\begin{ytableau}
            \yt{2}
        \end{ytableau}} &= 6 h_3 + 4 \sigma_3 \psi_0, &\hspace{15mm}{\bf S}_{\,\begin{ytableau}
            \yt{2}
        \end{ytableau}} &= {\bf P}_{1,1}^2 + h_3 \, {\bf P}_{2,1} +\sqrt{\frac{(1 + h_3 h_2 \psi_0)(1+ h_3 h_1 \psi_0)}{\psi_0(1 + h_1 h_2 \psi_0)}}\, {\bf P}_{2,2}
    \end{aligned}
\end{align}
Here we choose the normalization of 3-Schur polynomials such as the coefficient of basis monomial ${\bf P}_{1,1}^2$ is equal to one. With this choice of normalization the Cauchy identity \eqref{Cauchy id} for the level 2 has the following form:
\begin{equation}
    \sum_{|\pi| = 2} \frac{{\bf S}_{\pi}({\bf P}) \cdot {\bf S}_{\pi}({\bf P}^{\prime})}{|| {\bf S}_{\pi} ||^2} = \frac{1}{2} \left( \frac{{\bf P}_{1,1} \cdot {\bf P}^{\prime}_{1,1}}{||{\bf P}_{1,1}||^2} \right)^2 + \frac{{\bf P}_{2,1} \cdot {\bf P}^{\prime}_{2,1}}{||{\bf P}_{2,1}||^2} + \frac{{\bf P}_{2,2} \cdot {\bf P}^{\prime}_{2,2}}{||{\bf P}_{2,2}||^2}
\end{equation}
\begin{align}
\begin{aligned}
    || {\bf S}_{\, \begin{ytableau}
        \ \\
        \
    \end{ytableau}} ||^2 &= \frac{2 \psi_0\left(h_1-h_2\right) \left(h_1-h_3\right)}{(1+ h_2 h_3 \psi_0)}, &\hspace{25mm} ||{\bf P}_{1,1}||^2 &= 1\\
    || {\bf S}_{\, \begin{ytableau}
        \ & \
    \end{ytableau}} ||^2 &= \frac{2 \psi_0\left(h_2-h_1\right) \left(h_2-h_3\right)}{(1+ h_1 h_3 \psi_0)}, &\hspace{25mm} ||{\bf P}_{2,1}||^2 &= 2\psi_0 \\
    || {\bf S}_{\, \begin{ytableau}
        \yt{2}
    \end{ytableau}} ||^2 &= \frac{2 \psi_0\left(h_3-h_2\right) \left(h_3-h_1\right)}{(1+ h_2 h_1 \psi_0)}, &\hspace{25mm} ||{\bf P}_{2,2}||^2 &= -2\psi_0
\end{aligned}
\end{align}
The multiplication rule and Littlewood-Richardson coefficients have the following from:
\begin{equation}
    {\bf S}_{\, \begin{ytableau}
        \
    \end{ytableau}} \cdot {\bf S}_{\, \begin{ytableau}
        \
    \end{ytableau}} = C_{\, \begin{ytableau}
        \
    \end{ytableau}, \begin{ytableau}
        \ \\
        \ 
    \end{ytableau}} \, {\bf S}_{\, \begin{ytableau}
        \ \\
        \
    \end{ytableau}} + C_{\, \begin{ytableau}
        \
    \end{ytableau}, \begin{ytableau}
        \ & \
    \end{ytableau}} \, {\bf S}_{\, \begin{ytableau}
        \ & \
    \end{ytableau}} + C_{\, \begin{ytableau}
        \
    \end{ytableau}, \begin{ytableau}
        \yt{2}
    \end{ytableau}} \, {\bf S}_{\, \begin{ytableau}
        \yt{2}
    \end{ytableau}}
\end{equation}
\begin{align}
\begin{aligned}
    C_{\, \begin{ytableau}
        \
    \end{ytableau}, \begin{ytableau}
        \ \\
        \ 
    \end{ytableau}} =\frac{1+ h_2 h_3 \psi_0}{\psi_0\left(h_1-h_2\right) \left(h_1-h_3\right) } \\
    C_{\, \begin{ytableau}
        \
    \end{ytableau}, \begin{ytableau}
        \ & \
    \end{ytableau}} =\frac{1 + h_1 h_3 \psi_0}{\psi_0\left(h_2-h_1\right) \left(h_2-h_3\right)} \\
     C_{\, \begin{ytableau}
        \
    \end{ytableau}, \begin{ytableau}
        \yt{2}
    \end{ytableau}} = \frac{1 + h_1 h_2 \psi_0}{\psi_0\left(h_3-h_1\right) \left(h_3-h_2\right) }
\end{aligned}
\label{LR coeffocients 2 level}
\end{align}
Now we provide the form of the $\psi_3$ operator and 3-Schur polynomials for the cases $\psi_0 = 1$ and $\psi_0 = - \frac{1}{\sigma_2}$. We compare these two approaches and discuss the advantages/disadvantages.
\subsubsection{$\psi_0 = 1$}
In this case the formulas are lengthier and involved:
\begin{equation}
    \setstretch{2.5}
         \begin{array}{c||c|c|c}
             \psi_{3} &  \psi_{3}\cdot{\bf P}_{1,1}^2 & \psi_{3}\cdot{\bf P}_{2,1} & \psi_{3}\cdot{\bf P}_{2,2}\\
             \hline
             \hline
             {\bf P}_{1,1}^2 &  4 \sigma_3  & 6 & 0 \\
             \hline
             {\bf P}_{2,1} & 6 & 10 \sigma_3  & -6 \sqrt{ 1 + \sigma_2 + \sigma_3^2 }\\
             \hline
             {\bf P}_{2,2} & 0 & 6 \sqrt{ 1 + \sigma_2 + \sigma_3^2  } & -2 \sigma_3 \\
        \end{array}
\end{equation}

\begin{equation}
\setstretch{2.5}
         \begin{array}{c||c|c|c}
             \text{3-Schurs} & {\bf S}_{\,\begin{ytableau}
                 \ \\ \
             \end{ytableau}} & {\bf S}_{\,\begin{ytableau}
                 \ & \
             \end{ytableau}} & {\bf S}_{\,\begin{ytableau}
                 \yt{2}
             \end{ytableau}}\\
             \hline
             \hline
             {\bf P}_{1,1}^2 &  1 & 1 & 1 \\
             \hline
             {\bf P}_{2,1} & h_1 & h_2 & h_3 \\
             \hline
             {\bf P}_{2,2} & \sqrt{\frac{(1 + h_1 h_2 )(1+ h_1 h_3 )}{(1 + h_2 h_3)}} & \sqrt{\frac{(1 + h_1 h_2 )(1+ h_2 h_3 )}{(1 + h_1 h_3)}} & \sqrt{\frac{(1 + h_3 h_2 )(1+ h_1 h_3 )}{(1 + h_2 h_1)}} \\
        \end{array}
\end{equation} 
In the case $\psi_0 = 1$ there is nice reduction procedure to the $2d$ case of Jack and ordinary Schur polynomials. Namely, the $2d$ reduction condition reads \cite{Prochazka:2015deb}:
\begin{equation}
\label{2d reduction condition}
    h_3 + \sigma_3 \psi_0 = 0
\end{equation}
For the case $\psi_0=1$ this condition admits one parameter family of solutions, that corresponds to Jack polynomials:
\begin{equation}
\label{Jack case parameters}
    h_1 = -h, \hspace{15mm} h_2 = \frac{1}{h}, \hspace{15mm} h_3 = h - \frac{1}{h}
\end{equation}
where $h$ is a complex parameter and the case of ordinary Schur polynomials corresponds to $h=1$. We provide the explicit form of Jack and Schur polynonials to demostrate that the 3-Schur polynomials coincide with them up to overall normalization and identification of time-variables ${\bf P}_{a,i} \to p_a$:
\begin{align}
    \begin{aligned}
        \left. {\bf S}_{\,\begin{ytableau}
            \ \\ \
        \end{ytableau}}\right|_{\substack{h_1 = -h \\ h_2 = 1/h \\ h_3 = h-1/h}} &= {\bf P}_{1,1}^2 - h \, {\bf P}_{2,1}, &\hspace{10mm}
        J_{\, \begin{ytableau}
            \ \\
            \
        \end{ytableau}} &= p_1^2 - h \, p_2 , &\hspace{5mm} &\varpropto\, (h=1) &\hspace{5mm} S_{\, \begin{ytableau}
            \ \\
            \
        \end{ytableau}} &= \frac{1}{2}\left(p_1^2 - \, p_2 \right)\\
        \left. {\bf S}_{\,\begin{ytableau}
            \ & \
        \end{ytableau}}\right|_{\substack{h_1 = -h \\ h_2 = 1/h \\ h_3 = h-1/h}} &= {\bf P}_{1,1}^2 + \frac{1}{h} \, {\bf P}_{2,1}, &\hspace{10mm}
        J_{\, \begin{ytableau}
            \ & \
        \end{ytableau}} &= p_1^2 + \frac{1}{h} \, p_2 , &\hspace{5mm} &\varpropto \, (h=1) &\hspace{5mm} S_{\, \begin{ytableau}
            \ & \
        \end{ytableau}} &= \frac{1}{2}\left(p_1^2 +\, p_2 \right)\\
    \end{aligned}
\end{align}
In this limit 3-Schur polynomials of non-planar diagrams decouple, i.e. the corresponding Littlewood-Richardson coefficients \eqref{LR coeffocients 2 level} vanish:
\begin{equation}
    \left. C_{\, \begin{ytableau}
        \
    \end{ytableau}, \begin{ytableau}
        \yt{2}
    \end{ytableau}} \right|_{\substack{h_1 = -h \\ h_2 = 1/h \\ h_3 = h-1/h}} = 0
\end{equation}
In our normalization it also can be seen from the fact that their modules diverge and they disappear from the Cauchy identity:
\begin{equation}
    \left. \frac{1}{|| {\bf S}_{\, \begin{ytableau}
        \yt{2}
    \end{ytableau}} ||^2} \right|_{\substack{h_1 = -h \\ h_2 = 1/h \\ h_3 = h-1/h}} = 0
\end{equation}
\subsubsection{$\psi_0 = - \frac{1}{\sigma_2}$}
In this case the formulas for $\psi_3$ operator and 3-Schur polynomials are simpler and have the following from:
\begin{equation}
    \setstretch{2.5}
         \begin{array}{c||c|c|c}
             \psi_{3} &  \psi_{3}\cdot{\bf P}_{1,1}^2 & \psi_{3}\cdot{\bf P}_{2,1} & \psi_{3}\cdot{\bf P}_{2,2}\\
             \hline
             \hline
             {\bf P}_{1,1}^2 &  -4 \, \frac{\sigma_3}{\sigma_2}  & 6 & 0 \\
             \hline
             {\bf P}_{2,1} & -6\,\sigma_2 & -10 \, \frac{\sigma_3}{\sigma_2}  & -6 \, \frac{\sigma_3}{\sigma_2}\\
             \hline
             {\bf P}_{2,2} & 0 & 6 \, \frac{\sigma_3}{\sigma_2}  & 2\,\frac{\sigma_3}{\sigma_2}  \\
        \end{array}
\hspace{20mm}
         \begin{array}{c||c|c|c}
             \text{3-Schurs} & {\bf S}_{\,\begin{ytableau}
                 \ \\ \
             \end{ytableau}} & {\bf S}_{\,\begin{ytableau}
                 \ & \
             \end{ytableau}} & {\bf S}_{\,\begin{ytableau}
                 \yt{2}
             \end{ytableau}}\\
             \hline
             \hline
             {\bf P}_{1,1}^2 &  1 & 1 & 1 \\
             \hline
             {\bf P}_{2,1} & h_1 & h_2 & h_3 \\
             \hline
             {\bf P}_{2,2} & \frac{\sigma_3}{h_1^2} & \frac{\sigma_3}{h_2^2} & \frac{\sigma_3}{h_3^2} \\
        \end{array}
\end{equation}
\begin{align}
    \begin{aligned}
        \lambda_{\,\begin{ytableau}
            \ \\ \
        \end{ytableau}} &= 6 h_1 - 4  \frac{\sigma_3}{\sigma_2} , &\hspace{15mm} 
        {\bf S}_{\,\begin{ytableau}
            \ \\ \
        \end{ytableau}} &= {\bf P}_{1,1}^2 + h_1 \, {\bf P}_{2,1} + \frac{\sigma_3}{(h_1)^2} \, {\bf P}_{2,2} \\
        \lambda_{\,\begin{ytableau}
            \ & \
        \end{ytableau}} &= 6  h_2 - 4  \frac{\sigma_3}{\sigma_2}, &\hspace{15mm}{\bf S}_{\,\begin{ytableau}
            \ & \
        \end{ytableau}} &= {\bf P}_{1,1}^2 + h_2 \, {\bf P}_{2,1} + \frac{\sigma_3}{(h_2)^2} \, {\bf P}_{2,2} \\
        \lambda_{\,\begin{ytableau}
            \yt{2}
        \end{ytableau}} &= 6  h_3 - 4  \frac{\sigma_3}{\sigma_2}, &\hspace{15mm}{\bf S}_{\,\begin{ytableau}
            \yt{2}
        \end{ytableau}} &= {\bf P}_{1,1}^2 + h_3 \, {\bf P}_{2,1} + \frac{\sigma_3}{(h_3)^2} \, {\bf P}_{2,2}
    \end{aligned}
    \label{2 level 3Schurs}
\end{align}
Another interesting property of these formulas is that they can be represented in a universal way. Namely, any eigenvector of the $\psi_3$ operator on the level 2 have the following form:
\begin{equation}
    {\bf S}_{|\pi|=2} = {\bf P}_{1,1}^2 + u \cdot {\bf P}_{2,1} + \frac{\sigma_3}{u^2} \cdot {\bf P}_{2,2}
\end{equation}
provided that $u$ solves the characteristic equation:
\begin{equation}
\boxed{
    0 = \det( \psi_3 - 6 u + 4 \frac{\sigma_3}{\sigma_2} ) \ \ \varpropto \ \ u^3 + u \, \sigma_2 - \sigma_3 = (u-h_1)(u-h_2)(u-h_3)
    }
    \label{char eq 2}
\end{equation}
As one can see in this case significant simplifications occur. Further we fix the value of $\psi_0$:
\begin{equation}
\boxed{
    \psi_0 = - \frac{1}{\sigma_2}
}
\end{equation}
However, this value of $\psi_0$ has one disadvantage: there is no Jack polynomials in the $2d$ limit, only ordinary Schur polynomials. Namely, if one solves the $2d$ reduction condition for $\psi_0 = -\frac{1}{\sigma_2}$ one obtains the result $h_3 = 0$ that correspond to Schur polynomials. Recall, in the case of Jack polynomials $\sigma_3$ has a non-zero value \eqref{Jack case parameters}, while for $h_3 = 0$ is vanishes.

\bigskip

\subsection{3 level}
There are six basis monomials: ${\bf P}_{1,1}^3, {\bf P}_{2,1}{\bf P}_{1,1}, {\bf P}_{2,2}{\bf P}_{1,1}, {\bf P}_{3,1}, {\bf P}_{3,2}, {\bf P}_{3,3}$ and six $3d$ Young diagrams:
\begin{equation}
    \begin{ytableau} 
        \ & \ & \ \\ 
    \end{ytableau}, 
    \begin{ytableau} 
        \ & \ \\
        \ \\ 
    \end{ytableau},
    \begin{ytableau}
        \ \\
        \ \\ 
        \ \\
    \end{ytableau}, 
    \begin{ytableau}
        \yt{2} & \
    \end{ytableau}, 
    \begin{ytableau}
        \yt{2} \\
        \
    \end{ytableau}, 
    \begin{ytableau}
        \yt{3} 
    \end{ytableau}
\end{equation}

First, we fix the freedom of choice of new time-variables ${\bf P}_{3,i}$ we impose the following conditions:
\begin{align}
    \begin{aligned}
        Y_{1,1|2}^{1,2} &= 0, &\hspace{15mm} Y_{1,1|3}^{1,2} &= 0, &\hspace{15mm} Y_{1,2|1}^{1,2} &= 0, &\hspace{15mm} Y_{1,2|3}^{1,2} &= 0 \\
        X_{1,1|2}^{1,2} &= 0, &\hspace{15mm} X_{1,1|3}^{1,2} &= 0, &\hspace{15mm} X_{1,2|1}^{1,2} &= 0, &\hspace{15mm} X_{1,2|3}^{1,2} &= 0 \\
    \end{aligned}
\end{align}

Next we impose the Yangian relation:
\begin{equation}
    [\psi_2, \psi_3] \cdot {\bf P}_{2,1} = 0
\end{equation}

Combining it with the normalization rule \eqref{times normalization} we obtain:
\begin{align}
    Y_{1,1|1}^{1,2} &= 9, &\hspace{25mm} Y_{1,2|2}^{1,2} &= 12 \\
    X_{1,1|1}^{1,2} &= -6\sigma_2, &\hspace{25mm} X_{1,2|2}^{1,2} &= -3\sigma_2 \\
\end{align}

Then we use the following relations:
\begin{align}
    \begin{aligned}
        \left( \psi_3 - [e_2,f_1] \right) \cdot {\bf P}_{2,1} &= 0 &\hspace{10mm} &\Rightarrow &\hspace{10mm} U_{1,1}^{3} &= - 24 \, \frac{\sigma_3}{\sigma_2}, &\hspace{10mm} U_{2,1}^{3} &= - 9 \, \frac{\sigma_3}{\sigma_2},  \\ \left( \psi_3 - [e_2,f_1] \right) \cdot {\bf P}_{2,2} &= 0 &\hspace{10mm} &\Rightarrow &\hspace{10mm} U_{1,2}^{3} &= 24 \, \frac{\sigma_3}{\sigma_2}, &\hspace{10mm} U_{2,2}^{3} &= 3 \, \frac{\sigma_3}{\sigma_2},  
    \end{aligned}
\end{align}

Finally, we use the following relations \eqref{Yangian relations}:
\begin{equation}
    \text{YRel}_{0,1} \cdot {\bf P}_{1,1} =0, \hspace{25mm} \text{YRel}_{1,1} \cdot {\bf P}_{1,1} =0 
\end{equation}
and derive the set of constrains:
\begin{equation}
    U_{1,3}^{3} = 0, \hspace{15mm} U_{3,1}^{3} = 0, \hspace{15mm} U_{3,3}^{3} = 3 \, \frac{\sigma_3}{\sigma_2}, \hspace{15mm} U_{2,3}^3 \, U_{3,2}^3 = 27 \, \left(\frac{\sigma_3}{\sigma_2}\right)^2 \left( 4\, \frac{\sigma_2^3}{\sigma_3^2} - 1 \right)
\end{equation}

As at the second level the ratio $U_{2,3}^3 / U_{3,2}^3$ is not fixed from the Yanian relations and corresponds to the normalization of time $ {\bf P}_{3,3}$. Our choice of normalization reads:
\begin{equation}
    U_{2,3}^{3} = 9 \, \frac{\sigma_3}{\sigma_2}, \hspace{25mm} U_{3,2}^3 = 3 \, \frac{\sigma_3}{\sigma_2}\, \left( 4\, \frac{\sigma_2^3}{\sigma_3^2} - 1 \right)
\end{equation}

At this stage we completely fixed the action of the $\psi_3$ operator on the level three:
\begin{equation}
    \setstretch{2.5}
        \begin{array}{c||c|c|c|c|c|c}
            \psi_{3} & \psi_{3}\cdot{\bf P}_{1,1}^3 & \psi_{3}\cdot{\bf P}_{2,1} {\bf P}_{1,1} & \psi_{3}\cdot{\bf P}_{2,2} {\bf P}_{1,1} & \psi_{3}\cdot{\bf P}_{3,1} & \psi_{3}\cdot{\bf P}_{3,2} & \psi_{3}\cdot{\bf P}_{3,3} \\
            \hline
            \hline
            {\bf P}_{1,1}^3 & -6 \, \frac{\sigma_3}{\sigma_2} & 6 & 0 & 0 & 0 & 0 \\
            \hline
            {\bf P}_{2,1}{\bf P}_{1,1} & -18 \, \sigma_2 & -12\, \frac{\sigma_3}{\sigma_2} & 6\, \frac{\sigma_3}{\sigma_2} & 18 & 0 & 0 \\
            \hline
            {\bf P}_{2,2}{\bf P}_{1,1} & 0 & -6\, \frac{\sigma_3}{\sigma_2} & 0 & 0 & 24 & 0 \\
            \hline
            {\bf P}_{3,1} & 0 & -12 \, \sigma_2 & 0 & -24\, \frac{\sigma_3}{\sigma_2} & 24\, \frac{\sigma_3}{\sigma_2} & 0 \\
            \hline
            {\bf P}_{3,2} & 0 & 0 & -6 \, \sigma_2 & -9\, \frac{\sigma_3}{\sigma_2} & 3\, \frac{\sigma_3}{\sigma_2} & 9\, \frac{\sigma_3}{\sigma_2} \\
            \hline
            {\bf P}_{3,3} & 0 & 0 & 0 & 0 & 3 \, \frac{\sigma_3}{\sigma_2}\, \left( 4\, \frac{\sigma_2^3}{\sigma_3^2} - 1 \right) & 3\, \frac{\sigma_3}{\sigma_2} \\
        \end{array}
\end{equation}
We compute the 3-Schur polynomials as the eigenvectors:
\begin{equation}
    \setstretch{2.5}
        \begin{array}{c||c|c|c|c|c|c}
            \lambda_{\pi} & 18 h_1 - 6 \, \frac{\sigma_3}{\sigma_2} & 18 h_2 - 6 \, \frac{\sigma_3}{\sigma_2} & 18 h_3 - 6 \, \frac{\sigma_3}{\sigma_2} & 6 (h_1+h_2) - 6 \, \frac{\sigma_3}{\sigma_2} & 6 (h_1+h_3) - 6 \, \frac{\sigma_3}{\sigma_2} & 6 (h_2+h_3) - 6 \, \frac{\sigma_3}{\sigma_2} \\
            \text{3-Schurs} & {\bf S}_{\, \begin{ytableau}
                \ \\ 
                \ \\
                \
            \end{ytableau}} & 
            {\bf S}_{\, \begin{ytableau}
                \ & \ & \
            \end{ytableau}} & 
            {\bf S}_{\, \begin{ytableau}
                \yt{3}
            \end{ytableau}} & 
            {\bf S}_{\, \begin{ytableau}
                \ & \ \\
                \
            \end{ytableau}} & 
            {\bf S}_{\, \begin{ytableau}
                \yt{2} \\ 
                \
            \end{ytableau}} & 
            {\bf S}_{\, \begin{ytableau}
                \yt{2} & \ 
            \end{ytableau}} \\
            \hline
            \hline
            {\bf P}_{1,1}^3 & 1 & 1 & 1 & 1 & 1 & 1 \\
            \hline
            {\bf P}_{2,1}{\bf P}_{1,1} & 3 h_1  & 3 h_2 & 3 h_3 & -h_3 & -h_2 & -h_1 \\
            \hline
            {\bf P}_{2,2}{\bf P}_{1,1} & \frac{3\sigma_3}{(h_1)^2} & \frac{3\sigma_3}{(h_2)^2} & \frac{3\sigma_3}{(h_3)^2} & -\frac{h_3^2}{\sigma_3}\left( 3\sigma_2 + h_3^2 \right)& -\frac{h_2^2}{\sigma_3}\left( 3\sigma_2 + h_2^2 \right) & -\frac{h_1^2}{\sigma_3}\left( 3\sigma_2 + h_1^2 \right) \\
            \hline
            {\bf P}_{3,1} & 2 h_1^2 &  2 h_2^2 &  2 h_3^2 &  \frac{\sigma_3}{h_3} & \frac{\sigma_3}{h_2} & \frac{\sigma_3}{h_1} \\
            \hline
            {\bf P}_{3,2} & \frac{3\sigma_3}{2h_1} & \frac{3\sigma_3}{2h_2} & \frac{3\sigma_3}{2h_3} & \frac{h_3^3}{4\sigma_3}\left( 5\sigma_2 + 3h_3^2 \right) & \frac{h_2^3}{4\sigma_3}\left( 5\sigma_2 + 3h_2^2 \right) & \frac{h_1^3}{4\sigma_3}\left( 5\sigma_2 + 3h_1^2 \right) \\
            \hline
            {\bf P}_{3,3} & K(h_1) & K(h_2) & K(h_3) & h_3^2 \left( \frac{1}{4} -  \frac{\sigma_2^3}{\sigma_3^2} \right) & h_2^2 \left( \frac{1}{4} -  \frac{\sigma_2^3}{\sigma_3^2} \right) & h_1^2 \left( \frac{1}{4} -  \frac{\sigma_2^3}{\sigma_3^2} \right) \\
        \end{array}
\end{equation}
where we define new function:
\begin{equation}
    K(h_i) = \frac{\sigma_3}{2h_i^2}\left( 2\frac{\sigma_2^2}{\sigma_3} + \frac{\sigma_2}{h_i} + \frac{\sigma_3}{h_i^2}\right)
\end{equation}

Note that the eigenvalues of $\psi_3$ that are obtained in our algorithm nicely fit \eqref{eigenvalues}. There is another important comment in order. The resulting 3-Schur polynomials have nice properties with respect to transposition of the $3d$ Young diagram in various planes. In particular, transposing the diagram in the plane with $i$-th and $j$-th axis one should swap $h_i$ and $h_j$ in the corresponding 3-Schur polynomials. For example:
\begin{align}
     {\bf S}_{\, \begin{ytableau}
        \ \\
        \ \\
        \
    \end{ytableau}} \Big|_{h_1 \leftrightarrow h_2}= {\bf S}_{\, \begin{ytableau}
        \ & \ & \
    \end{ytableau}}, \hspace{10mm}
     {\bf S}_{\, \begin{ytableau}
        \ & \ \\
        \
    \end{ytableau}} \Big|_{h_1 \leftrightarrow h_2}= {\bf S}_{\, \begin{ytableau}
        \ & \ \\
        \
    \end{ytableau}}, \hspace{10mm}
     {\bf S}_{\, \begin{ytableau}
        \ \\
        \ \\
        \
    \end{ytableau}} \Big|_{h_1 \leftrightarrow h_3}= {\bf S}_{\, \begin{ytableau}
        \yt{3}
    \end{ytableau}}
\end{align}
Note that this peculiar property of diagram transposition is not required in our algorithm and appears automatically. One can check that 3-Schur polynomials on the level 2 \eqref{2 level 3Schurs} also obey this transposition property. 

The universal formulas for 3-Schurs on the level 3 in terms of parameter $u$ have the following form:
\begin{align}
\begin{aligned}
    {\bf S}_{|\pi|=3} &= {\bf P}_{1,1}^3 + u \cdot {\bf P}_{2,1} {\bf P}_{1,1} + \left( \frac{9 \sigma _2^2}{4 \sigma _3}+\frac{\sigma _2 \left(9 \sigma _3+u^3\right)}{4 \sigma _3 u}+u \right) \cdot  {\bf P}_{2,2} {\bf P}_{1,1} + \\
    &+\left( \frac{u^3+\sigma_2 u -3 \sigma_3}{4 u} \right) \cdot {\bf P}_{3,1} + 
    \left( \frac{\sigma _2 u^4+3 \sigma _3 \left(u^3-3 \sigma _3\right)+9 \sigma _2^2 u^2}{16 \sigma _3 u}\right) \cdot {\bf P}_{3,2} + \\
    &+ \left( \frac{\sigma _2^3 \left(72 \sigma _3+26 u^3\right)+3 \sigma _3 \sigma _2 u \left(u^3-2 \sigma _3\right)+3 \sigma _3^2 \left(u^3-3 \sigma _3\right)+\sigma _2^2 \left(2 u^5+5 \sigma _3 u^2\right)+72 \sigma _2^4 u}{48 \sigma _3^2 u}\right) \cdot {\bf P}_{3,3}
\end{aligned}
\end{align}
where $u$ one of the solutions of the characteristic equation:
\begin{align}
\boxed{
    \begin{aligned}
         0 &= \det( \psi_3 - 6 u + 6 \frac{\sigma_3}{\sigma_2} ) \ \ \varpropto \ \ \left( u^3+9 \sigma_2 u-27 \sigma _3 \right)\left(u^3+\sigma _2 u+\sigma _3\right) =\\
         &=(u-3h_1)(u-3h_2)(u-3h_3)(u+h_1)(u+h_2)(u+h_3)
    \end{aligned}
    }
    \label{char eq 3}
\end{align}
Note that the eigenvalues $6u - 4\frac{\sigma_3}{\sigma_2}$ are in the agreement with formula \eqref{eigenvalues}.
The above 3-Schur polynomials on the level 3 obey the Cauchy identity \eqref{Cauchy id}:
\begin{align}
    \begin{aligned}
        \sum_{|\pi| = 3} \frac{{\bf S}_{\pi}({\bf P}) \cdot {\bf S}_{\pi}({\bf P}^{\prime})}{|| {\bf S}_{\pi} ||^2} &= \frac{1}{6} \left( \frac{{\bf P}_{1,1} \cdot {\bf P}^{\prime}_{1,1}}{||{\bf P}_{1,1}||^2} \right)^3 + \frac{{\bf P}_{2,1}{\bf P}_{1,1} \cdot {\bf P}^{\prime}_{2,1}{\bf P}^{\prime}_{1,1}}{||{\bf P}_{2,1}||^2||{\bf P}_{1,1}||^2} + \frac{{\bf P}_{2,2}{\bf P}_{1,1} \cdot {\bf P}^{\prime}_{2,2}{\bf P}^{\prime}_{1,1}}{||{\bf P}_{2,2}||^2||{\bf P}_{1,1}||^2}+\\
        &+\frac{{\bf P}_{3,1} \cdot {\bf P}^{\prime}_{3,1}}{||{\bf P}_{3,1}||^2} + \frac{{\bf P}_{3,2} \cdot {\bf P}^{\prime}_{3,2}}{||{\bf P}_{3,2}||^2} + 
        \frac{{\bf P}_{3,3} \cdot {\bf P}^{\prime}_{3,3}}{||{\bf P}_{3,3}||^2}
    \end{aligned}
\end{align}

\begin{align}
    \begin{aligned}
        ||{\bf S}_{\, \begin{ytableau}
        \ \\
        \ \\
        \
        \end{ytableau}}||^2 &= \frac{6 \left(h_1-h_2\right) \left(2 h_1-h_2\right) \left(h_1-h_3\right) \left(2 h_1-h_3\right)}{ h_1 \left(\sigma_3 -2 h_1 \sigma _2\right)},&\hspace{25mm} ||{\bf P}_{1,1}||^2 &= 1\\
        ||{\bf S}_{\, \begin{ytableau}
        \ & \ & \ 
        \end{ytableau}} ||^2&= \frac{6 \left(h_2-h_1\right) \left(2 h_2-h_1\right) \left(h_2-h_3\right) \left(2 h_2-h_3\right)}{ h_2 \left(\sigma_3 -2 h_2 \sigma _2\right)}, &\hspace{25mm} ||{\bf P}_{2,1}||^2 &= -\frac{2}{\sigma_2} \\
        ||{\bf S}_{\, \begin{ytableau}
        \yt{3}
        \end{ytableau}}||^2 &= \frac{6 \left(h_3-h_2\right) \left(2 h_3-h_2\right) \left(h_3-h_1\right) \left(2 h_3-h_1\right)}{ h_3 \left(\sigma_3 -2 h_3 \sigma _2\right)}, &\hspace{25mm} ||{\bf P}_{2,2}||^2 &= \frac{2}{\sigma_2}\\
        ||{\bf S}_{\, \begin{ytableau}
        \ & \ \\
        \ 
        \end{ytableau}}||^2 &= \frac{\left(h_2-2 h_1\right) \left(h_1-2h_2\right) \left(h_3-h_2\right) \left(h_3-h_1\right)}{h_1^2 h_2^2}, &\hspace{25mm} ||{\bf P}_{3,1}||^2 &= \frac{3}{\sigma_2^2}\\
        ||{\bf S}_{\, \begin{ytableau}
        \yt{2} \\
        \  
        \end{ytableau}}||^2&= \frac{\left(h_3-2 h_1\right) \left(h_1-2h_3\right) \left(h_2-h_3\right) \left(h_2-h_1\right)}{h_1^2 h_3^2}, &\hspace{25mm} ||{\bf P}_{3,2}||^2 &= -\frac{8}{\sigma_2^2} \\
        ||{\bf S}_{\, \begin{ytableau}
        \yt{2} & \ 
        \end{ytableau}}||^2&= \frac{\left(h_2-2 h_3\right) \left(h_3-2h_2\right) \left(h_1-h_2\right) \left(h_1-h_3\right)}{h_3^2 h_2^2}, &\hspace{25mm} ||{\bf P}_{3,3}||^2 &= \frac{24}{\sigma _2^2 \left(1-\frac{4 \sigma _2^3}{\sigma _3^2}\right)} \\
    \end{aligned}
\end{align}
Finally, it can be checked that the above defined 3-Schur functions reproduce ordinary Schur functions in $2d$ limit:
\begin{align}
    \begin{aligned}
        {\bf S}_{\, \begin{ytableau}
        \ \\
        \ \\
        \ 
    \end{ytableau}} \Big|_{\substack{h_1 = -1 \\ h_2 = 1 \\ h_3 = 0}} &= p_1^3-3 p_2 p_1+2 p_3, &\hspace{15mm}
    {\bf S}_{\, \begin{ytableau}
        \ & \ & \  
    \end{ytableau}} \Big|_{\substack{h_1 = -1 \\ h_2 = 1 \\ h_3 = 0}} &= p_1^3+3 p_2 p_1+2 p_3 \\
     {\bf S}_{\, \begin{ytableau}
        \ & \ \\
        \ 
    \end{ytableau}} \Big|_{\substack{h_1 = -1 \\ h_2 = 1 \\ h_3 = 0}} &= p_1^3-p_3, &\hspace{15mm} &
    \end{aligned}
\end{align}

\bigskip

\subsection{4 level}
On the level 4 there are 13 basis monomials: ${\bf P}_{4,1}, \ {\bf P}_{4,2}, \ {\bf P}_{4,3}, \ {\bf P}_{4,4}, \ ({\bf P}_{2,1})^2, \ {\bf P}_{2,1}{\bf P}_{2,2}, \ ({\bf P}_{2,2})^2$, \\$ {\bf P}_{3,1}{\bf P}_{1,1}, \ {\bf P}_{3,2}{\bf P}_{1,1}, \ {\bf P}_{3,3}{\bf P}_{1,1}, \ {\bf P}_{2,1} ({\bf P}_{1,1})^2, \ {\bf P}_{2,2} ({\bf P}_{1,1})^2, \ ({\bf P}_{1,1})^4$ and $3d$ Young diagrams:
\begin{equation}
     \begin{ytableau}
            \ & \ & \ & \ 
        \end{ytableau}, \begin{ytableau}
            \ & \ & \ \\
            \
        \end{ytableau},  \begin{ytableau}
            \ & \ \\
            \ & \ 
        \end{ytableau},  \begin{ytableau}
            \ & \ \\
            \ \\
            \
        \end{ytableau}, \begin{ytableau}
            \ \\
            \ \\
            \ \\
            \
        \end{ytableau}, 
        \begin{ytableau}
            \yt{2} & \ & \ \\
        \end{ytableau}, 
        \begin{ytableau}
            \yt{2} \\
            \ \\
            \ \\
        \end{ytableau}, 
        \begin{ytableau}
            \yt{2} & \yt{2} 
        \end{ytableau},
        \begin{ytableau}
            \yt{2} \\
            \yt{2}
        \end{ytableau},
        \begin{ytableau}
            \yt{3} & \ 
        \end{ytableau},
        \begin{ytableau}
            \yt{3} \\
            \
        \end{ytableau},
        \begin{ytableau}
            \yt{2} & \ \\
            \ 
        \end{ytableau},
        \begin{ytableau}
            \yt{4}
        \end{ytableau} 
\end{equation}
Firstly, we fix a basis in subspace of times ${\bf P}_{4,i}$:
\begin{align}
    Y_{1,1|2}^{1,3} &=0, &\ \ Y_{1,1|3}^{1,3} &=0, &\ \ Y_{1,1|4}^{1,3} &=0, &\ \ Y_{1,2|1}^{1,3} &=0, &\ \ Y_{1,2|3}^{1,3} &=0, &\ \ Y_{1,2|4}^{1,3} &=0 \\
    X_{1,1|2}^{1,3} &=0, &\ \ X_{1,1|3}^{1,3} &=0, &\ \ X_{1,1|4}^{1,3} &=0, &\ \ X_{1,2|1}^{1,3} &=0, &\ \ X_{1,2|3}^{1,3} &=0, &\ \ X_{1,2|4}^{1,3} &=0 \\
    Y_{1,3|1}^{1,3} &=0, &\ \ X_{1,3|1}^{1,3} &=0, &\ \ Y_{1,3|2}^{1,3} &=0, &\ \ X_{1,3|2}^{1,3} &=0, &\ \ Y_{1,3|4}^{1,3} &=0, &\ \ X_{1,3|4}^{1,3} &=0 
\end{align}
The number of these constrains exceed the dimension of $GL(4)$ that changes the basis in the subspace of time-variables ${\bf P}_{4,i}$. However, not all of the constrains are independent and the above set is consistent. 

Next we list the Yangian relations and their consequences for the ansatz coefficents:
\begin{equation}
    \left( [\psi_2, \psi_3]\right) \cdot {\bf P}_{3,2} = 0
\end{equation}
\begin{align}
\begin{aligned}
     X_{1,1|1}^{1,3} &= -9 \, \sigma_2, &\hspace{15mm} Y_{1,1|1}^{1,3} &= 12, \\
     X_{1,2|2}^{1,3} &= -6 \, \sigma_2, &\hspace{15mm} Y_{1,2|2}^{1,3} &= 15, \\
     X_{1,3|3}^{1,3} &= -3 \, \sigma_2, &\hspace{15mm} Y_{1,3|3}^{1,3} &= 18 \\
\end{aligned}
\end{align}
We impose the following condition for arbitrary $\alpha_i$:
\begin{equation}
    \left( \psi_3 - [e_2,f_1] \right) \cdot \left( \alpha_1 {\bf P}_{2,1} {\bf P}_{1,1} + \alpha_2 {\bf P}_{2,2} {\bf P}_{1,1} + \alpha_3 {\bf P}_{3,1} +\alpha_4 {\bf P}_{3,2} + \alpha_5 {\bf P}_{3,3} \right)  = 0
\end{equation}
\begin{align}
\begin{aligned}
     X_{1,1|1}^{2,2} &= -12 \, \sigma_2, &\hspace{15mm} X_{1,1|2}^{2,2} &= 0, &\hspace{15mm} X_{1,1|3}^{2,2} &= 0, \\
     X_{1,2|1}^{2,2} &= 0, &\hspace{15mm} X_{1,2|2}^{2,2} &= -\frac{18}{5} \, \sigma_2, &\hspace{15mm} X_{1,2|3}^{2,2} &= 0,\\
     Y_{1,1|1}^{2,2} &= 12, &\hspace{15mm} Y_{1,1|2}^{2,2} &= 0, &\hspace{15mm}  Y_{1,1|3}^{2,2} &= 0, \\
     Y_{1,2|1}^{2,2} &= 0, &\hspace{15mm} Y_{1,2|2}^{2,2} &= 18, &\hspace{15mm}  Y_{1,2|3}^{2,2} &= 0,\\
     U_{1,1}^{4} &= -44 \, \frac{\sigma_3}{\sigma_2}, &\hspace{15mm} U_{1,2}^{4} &= 60 \, \frac{\sigma_3}{\sigma_2},   &\hspace{15mm} U_{1,3}^{4} &= 0, \\
     U_{2,1}^{4} &= -12 \, \frac{\sigma_3}{\sigma_2}, &\hspace{15mm} U_{2,2}^{4} &= 4 \, \frac{\sigma_3}{\sigma_2}, &\hspace{15mm} U_{2,3}^{4} &= \frac{144\sigma_3}{5\sigma_2}, \\
     U_{3,1}^{4} &= 0, &\hspace{15mm} U_{3,2}^{4} &= 4 \, \frac{\sigma_3}{\sigma_2}\, \left( 4\, \frac{\sigma_2^3}{\sigma_3^2} - 1 \right), &\hspace{15mm} U_{3,3}^{4} &= 4 \, \frac{\sigma_3}{\sigma_2}. \\
\end{aligned}
\end{align}
\begin{align}
    \begin{aligned}
         \text{YRel}_{0,0} \cdot \left( \alpha_1 {\bf P}_{1,1}^2 + \alpha_2 {\bf P}_{2,1} \alpha_3 {\bf P}_{2,2} \right) & =0 \\
         \text{YRel}_{1,1} \cdot \left( \alpha_1 {\bf P}_{1,1}^2 + \alpha_2 {\bf P}_{2,1} \alpha_3 {\bf P}_{2,2} \right) & =0 \\
    \end{aligned}
\end{align}
\begin{align}
    \begin{aligned}
        X_{1,1|4}^{2,2} &= 0 &\hspace{10mm}  X_{2,2|1}^{2,2} &= 0 &\hspace{10mm} 
        X_{2,2|2}^{2,2} &= 0 &\hspace{10mm} 
        X_{2,2|4}^{2,2} &= 0 \\
        Y_{1,1|4}^{2,2} &= 0 &\hspace{10mm}  Y_{2,2|1}^{2,2} &= 0 &\hspace{10mm} 
        Y_{2,2|2}^{2,2} &= 0 &\hspace{10mm} 
        Y_{2,2|4}^{2,2} &= 0 \\
        U_{1,4}^{4} &= 0 &\hspace{10mm}   U_{2,4}^{4} &= 0 &\hspace{10mm} 
        U_{4,1}^{4} &= 0 &\hspace{10mm} 
        U_{4,2}^{4} &= 0 \\
    \end{aligned}
\end{align}
\begin{align}
    \begin{aligned}
        X_{1,2|4}^{2,2} Y_{1,2|4}^{2,2} = -\frac{396}{5} \sigma_2 \hspace{10mm} U_{4,4}^{4} = 4 \frac{\sigma_3}{\sigma_2} \hspace{10mm} U_{3,4}^{4} = \tau \cdot Y_{1,2|4}^{2,2} \cdot \frac{\sigma_3}{\sigma_2}
    \end{aligned}
\end{align}
\begin{equation}
    X_{2,2|3}^{2,2} = \frac{2}{5} \sigma _2 \left(\frac{16 \sigma _2^3}{\sigma _3^2}-33 \tau -4\right)
\end{equation}
\begin{equation}
    Y_{2,2|3}^{2,2}= B \cdot 72 \left(12 \tau +1-14 \frac{\sigma _2^3}{\sigma_3^2}\right)
\end{equation}
\begin{equation}
    U_{4,3}^{4}=-B \cdot \frac{432  \left(\sigma _3^2 \left(5 \sigma _3^2-6 \sigma _2^3 (44 \tau +17)\right)+128 \sigma _2^6\right)}{5
   \sigma _2 \sigma_3^3Y_{1,2|4}^{2,2}}
\end{equation}
Where parameter $\tau$ should obey the following equation:
\begin{equation}
    \sigma _3^4 \left(1287 \tau ^2-93 \tau +13\right)-\sigma _3^2 \sigma _2^3 (3693 \tau +424)+1488 \sigma _2^6 = 0
\end{equation}
and we defined a new function:
\begin{equation}
    B = \frac{\sigma_3^4}{2 \sigma _3^2 \sigma _2^3 (33 \tau +4)+15
   \sigma _3^4 \tau -32 \sigma _2^6}
\end{equation}
Now we are able to compute 3-Schur polynomials. For this purpose we define the following functions:
\begin{equation}
    K(h_i) = \frac{\sigma_3}{2h_i^2}\left( 2\frac{\sigma_2^2}{\sigma_3} + \frac{\sigma_2}{h_i} + \frac{\sigma_3}{h_i^2}\right)
\end{equation}
\begin{equation}
    L(h_i) = -B\cdot\frac{3   }{h_i^7 \sigma_3^2}\left(-6 h_i^3 \sigma _3^2 \sigma _2^3 (\tau -2)+6 h_i^2 \sigma _3^3 \sigma _2^2 (2 \tau +1)+h_i
   \sigma _3^4 \sigma _2 (9 \tau +2)+8 h_i^5 \sigma _2^5+12 h_i^4 \sigma _3 \sigma _2^4-15 \sigma _3^5 \tau
   \right)
\end{equation}
\begin{align}
\begin{aligned}
    M(h_i) &= B \cdot \frac{36}{5\sigma_3^3 h_i^9 Y_{1,2|4}^{2,2}
   }\Big(  2 h_i^5 \sigma _3^2 \sigma _2^5 (33 \tau -46)-2 h_i^4 \sigma _3^3 \sigma _2^4 (66 \tau
   +13) -3 h_i^3 \sigma _3^4 \sigma _2^3 (33 \tau +4)+5 h_i^2 \sigma _3^5 \sigma _2^2 (33 \tau +7)+\\
   &+32 h_i^7 \sigma_2^7-32 h_i^6 \sigma _3 \sigma _2^6+10 h_i \sigma _3^6 \sigma _2+5 \sigma _3^7 \Big)
\end{aligned}
\end{align}
\begin{align}
\begin{aligned}
    \lambda_{\, \begin{ytableau}
        \ \\
        \ \\
        \ \\
        \
    \end{ytableau}} &= 36 h_1 - 8 \frac{\sigma_3}{\sigma_2} \\
    {\bf S}_{\, \begin{ytableau}
        \ \\
        \ \\
        \ \\
        \
    \end{ytableau}} &= ({\bf P}_{1,1})^4 + 6h_1 \cdot {\bf P}_{2,1} {\bf P}_{1,1}^2 + \frac{6\sigma_3}{h_1^2} \cdot {\bf P}_{2,2} {\bf P}_{1,1}^2 + 8h_1^2 \cdot {\bf P}_{3,1} {\bf P}_{1,1} + \frac{6\sigma_3}{h_1} \cdot {\bf P}_{3,2} {\bf P}_{1,1} +4 K(h_1) \cdot {\bf P}_{3,3} {\bf P}_{1,1} +\\
    & +3 h_1^2 \cdot {\bf P}_{2,1}^2+\frac{6 \sigma _3}{h_1} \cdot {\bf P}_{2,1}{\bf P}_{2,2}+ L(h_1) \cdot {\bf P}_{2,2}^2 + 6 h_1^3 \cdot {\bf P}_{4,1} + \frac{18 \sigma _3}{5} \cdot {\bf P}_{4,2}+2 h_1 K(h_1) \cdot {\bf P}_{4,3} + M(h_1) \cdot {\bf P}_{4,4}
\end{aligned}
\end{align}

\begin{align}
\begin{aligned}
    \lambda_{\, \begin{ytableau}
        \ & \ \\
        \ \\
        \
    \end{ytableau}} &= 18 h_1 + 6h_2 - 8 \frac{\sigma_3}{\sigma_2} \\
    {\bf S}_{\, \begin{ytableau}
        \ & \ \\
        \ \\
        \ 
    \end{ytableau}} &= ({\bf P}_{1,1})^4 + (3 h_1+h_2) \cdot {\bf P}_{2,1} {\bf P}_{1,1}^2 + \frac{h_3^2 \left(2 h_2^2-h_1 h_2-3 \sigma _2\right)}{\sigma _3} \cdot {\bf P}_{2,2} {\bf P}_{1,1}^2 + (-2 h_1 h_3) \cdot {\bf P}_{3,1} {\bf P}_{1,1} + \\
    &+\frac{h_3^2 \left(h_1 \left(4 h_2+3 h_3\right) \sigma _2-3 h_2 \sigma _3\right)}{2 h_1 \sigma _3} \cdot {\bf P}_{3,2} {\bf P}_{1,1} -\frac{h_3^2+4 \sigma _2}{h_2^2} \cdot K(h_1) \cdot {\bf P}_{3,3} {\bf P}_{1,1} +\\
    & +h_1 h_2 \cdot {\bf P}_{2,1}^2+\frac{h_3^2 \left(4 h_2 \sigma _2+3 h_3 \sigma _2+2 h_2^3\right)}{\sigma _3} \cdot {\bf P}_{2,1}{\bf P}_{2,2}+ \frac{h_1 h_3^2 \left(\sigma_3-3 h_1 \sigma _2\right)}{3 \sigma _3^2} \cdot L(h_1)  \cdot {\bf P}_{2,2}^2 + \\
    &+2 h_1^2 h_2\cdot {\bf P}_{4,1} + \frac{h_3 \left(9 h_3 \sigma _2-2 h_2 \sigma _2+6 h_3^3\right)}{5 h_2} \cdot {\bf P}_{4,2}+\frac{h_1^2 h_3^2 \left(5 h_3 \sigma _2+2 h_2 \sigma _2+2 h_3^3\right)}{3 \sigma _3^2}\cdot  K(h_1) \cdot {\bf P}_{4,3} + \\
    &+ \frac{h_1 h_3^2 \left(\sigma _3-3 h_1 \sigma _2\right)}{3 \sigma _3^2} \cdot M(h_1) \cdot {\bf P}_{4,4}
\end{aligned}
\end{align}

\begin{align}
\begin{aligned}
    \lambda_{\, \begin{ytableau}
        \ & \ \\
        \ & \
    \end{ytableau}} &= 12 (h_1+h_2) - 8 \frac{\sigma_3}{\sigma_2} \\
    {\bf S}_{\, \begin{ytableau}
        \ & \ \\
        \ & \
    \end{ytableau}} &= ({\bf P}_{1,1})^4 + (-2 h_3)\cdot {\bf P}_{2,1} {\bf P}_{1,1}^2 -\frac{2 h_3^2 \left(h_3^2+3 \sigma _2\right)}{\sigma _3} \cdot {\bf P}_{2,2} {\bf P}_{1,1}^2 + \frac{4 \sigma _3}{h_3} \cdot {\bf P}_{3,1} {\bf P}_{1,1} + \frac{5 h_3^3 \sigma _2+3 h_3^5}{\sigma _3} \cdot {\bf P}_{3,2} {\bf P}_{1,1} +\\
    +&h_3^2 \left(1-\frac{4 \sigma _2^3}{\sigma _3^2}\right) \cdot {\bf P}_{3,3} {\bf P}_{1,1}  -(2 h_3^2+3 \sigma _2)\cdot {\bf P}_{2,1}^2-4 h_3^2 \cdot {\bf P}_{2,1}{\bf P}_{2,2}+\\
    &-B\cdot\frac{h_3^2}{\sigma_3^4} \Big( h_3^{10} \sigma _2 (78 \tau -1)+2 h_3^8 \sigma _2^2 (48 \tau -1)+14 h_3^6 \sigma _2^3 (12 \tau +1)+ 2 h_3^4 \sigma _2^4 (105 \tau +17)+\\
    &+15 h_3^2 \sigma _2^5 (6 \tau +1)+30 h_3^{12} \tau -60 \sigma _2^6 \Big) \cdot {\bf P}_{2,2}^2 -\sigma _3 \cdot {\bf P}_{4,1} -\frac{5 h_3^4 \sigma _2+3 h_3^6}{5 \sigma _3} \cdot {\bf P}_{4,2}+ \frac{h_3^3 \left(4 \sigma _2^3-\sigma _3^2\right)}{6 \sigma _3^2} \cdot {\bf P}_{4,3}\, -\\
    &-B\cdot\frac{6 h_3^2}{5 \sigma _3^5 Y_{1,2|4}^{2,2}} \Big(5 h_3^{12} \sigma _2^2 (132 \tau +7)+4 h_3^{10} \sigma _2^3 (429 \tau -23)+h_3^8 \sigma _2^4 (2112 \tau -229)+\\
    &+12 h_3^6 \sigma _2^5 (308 \tau +9)+h_3^4 \sigma _2^6 (4620 \tau +893)+220 h_3^2 \sigma _2^7 (9 \tau +1)+20 h_3^{14} \sigma _2+5
   h_3^{16}-960 \sigma _2^8\Big) \cdot {\bf P}_{4,4}
\end{aligned}
\end{align}

\begin{align}
\begin{aligned}
    \lambda_{\, \begin{ytableau}
        \yt{2} & \ \\
        \ 
    \end{ytableau}} &=  - 8 \frac{\sigma_3}{\sigma_2} \\
    {\bf S}_{\, \begin{ytableau}
        \yt{2} & \ \\
        \ 
    \end{ytableau}} &= ({\bf P}_{1,1})^4 + 0 \cdot {\bf P}_{2,1} {\bf P}_{1,1}^2 + \frac{24 \sigma _2^2}{7 \sigma _3} \cdot {\bf P}_{2,2} {\bf P}_{1,1}^2 + \frac{4 \sigma _2}{7} \cdot {\bf P}_{3,1} {\bf P}_{1,1} -\frac{4 \sigma _2}{7} \cdot {\bf P}_{3,2} {\bf P}_{1,1} +\frac{6\sigma _2}{7}  \left(\frac{4 \sigma _2^3}{\sigma _3^2}-1\right) \cdot {\bf P}_{3,3} {\bf P}_{1,1} +\\
    & +\frac{3 \sigma _2}{7} \cdot {\bf P}_{2,1}^2-\frac{8 \sigma _2}{7} \cdot {\bf P}_{2,1}{\bf P}_{2,2} +B\cdot\frac{1}{7 \sigma_3^4}\left(2 \sigma_3^2 \sigma _2^4 (36 \tau +47)-\sigma _3^4 \sigma _2 (144 \tau +7)-264 \sigma _2^7\right) \cdot {\bf P}_{2,2}^2 +\\
    &+\sigma _3 \cdot {\bf P}_{4,1} + \frac{3 \left(4 \sigma _2^3+7 \sigma _3^2\right)}{35 \sigma _3} \cdot {\bf P}_{4,2}+\frac{\sigma _3^2-4 \sigma _2^3}{6 \sigma _3} \cdot {\bf P}_{4,3} +\\
    &+B\cdot \frac{6}{35 \sigma _3^5 Y_{1,2|4}^{2,2} } \left(24 \sigma _3^2 \sigma _2^6 (66 \tau +127)-2 \sigma _3^4 \sigma _2^3 (1584 \tau +427)-768 \sigma _2^9+35 \sigma _3^6\right) \cdot {\bf P}_{4,4}
\end{aligned}
\end{align}

The characteristic equation for the $u$ parameter:
\begin{align}
\boxed{
    \begin{aligned}
        0 &= \det( \psi_3 - 6 u + 8 \frac{\sigma_3}{\sigma_2} ) \ \ \varpropto \\
        & \varpropto \ \ u \left( u^3+36 \sigma_2 u -216 \sigma_3 \right) \left(u^3+4 \sigma _2 u + 8 \sigma_3\right) \left(u^6+14 \sigma _2 u^4-20 \sigma _3 u^3+49 \sigma _2^2 u^2-140 \sigma _2 \sigma _3 u+36 \sigma _2^3+343 \sigma _3^2 \right) = \\
        & =  u(u-6h_1)(u-6h_2)(u-6h_3)(u+2h_3)(u+2h_2)(u+2h_1)\times \\
        &\times(u-3h_1-h_2)(u-3h_2-h_1)(u-3h_2 -h3)(u-3h_3-h_2)(u-3h_1 -h_3)(u-3h_3-h_1)
    \end{aligned}
    }
    \label{char eq 4}
\end{align}

Despite the formulas for cut-and-join operator and 3-Schur functions are complicated the values of the eigenvalues are nice and obey \eqref{eigenvalues}. Note that the most sophisticated answers appear for the coefficients in front of time variables ${\bf P}_{2,2}^2$ and ${\bf P}_{4,4}$, while the other coefficients are the way simpler. The Cauchy identity reveals another flaw in our choice of basis in the time variables, especially in the sector ${\bf P}_{2,2}^2$, ${\bf P}_{4,4}$.

\begin{align}
    \begin{aligned}
        ||{\bf S}_{\, \begin{ytableau}
        \ \\
        \ \\
        \ \\
        \
        \end{ytableau}}||^2 &= \frac{24 \left(h_1-h_2\right) \left(2 h_1-h_2\right) \left(3 h_1-h_2\right) \left(h_1-h_3\right) \left(2 h_1-h_3\right) \left(3 h_1-h_3\right)}{\left(2 h_1 \sigma _2-\sigma _3\right) \left(3 h_1 \sigma _2-\sigma _3\right)} \\
        ||{\bf S}_{\, \begin{ytableau}
        \ & \ \\
        \ \\
        \ 
        \end{ytableau}}||^2 &=\frac{4 h_1\left(h_1-h_2\right)^2 \left(h_2-3 h_1\right) \left(h_2-h_3\right) h_3^2 \left(h_3-2 h_1\right) \left(h_1-h_3\right)}{\sigma _3^2 \left(2 \sigma _2 h_1 - \sigma _3 \right)} \\
        ||{\bf S}_{\, \begin{ytableau}
        \ & \ \\
        \ & \ 
        \end{ytableau}}||^2&=\frac{8 \left(h_1-2 h_2\right) \left(h_1-h_2\right){}^2 \left(2 h_1-h_2\right) \left(h_1-h_3\right) \left(h_2-h_3\right) h_3^2}{\sigma _3^2 \left(h_3^2+2 \sigma _2\right)} \\
        ||{\bf S}_{\, \begin{ytableau}
        \yt{2} & \ \\
        \ 
        \end{ytableau}}||^2 &=\frac{8 \left(2h_2- h_1\right) \left(2 h_1-h_2\right) \left(h_1-2 h_3\right) \left(h_2-2 h_3\right) \left(2 h_1-h_3\right) \left(2 h_2-h_3\right)}{49 \sigma _3^2}
    \end{aligned} 
\end{align}
The Cauchy identity \eqref{Cauchy id} is slightly violated in the mentioned sector:
\begin{align}
    \begin{aligned}
        \sum_{|\pi| = 4} \frac{{\bf S}_{\pi}({\bf P}) \cdot {\bf S}_{\pi}({\bf P}^{\prime})}{|| {\bf S}_{\pi} ||^2} &= \frac{1}{24} \left( \frac{{\bf P}_{1,1} \cdot {\bf P}^{\prime}_{1,1}}{||{\bf P}_{1,1}||^2} \right)^4 + \frac{1}{2}\frac{{\bf P}_{2,1}({\bf P}_{1,1})^2 \cdot {\bf P}^{\prime}_{2,1}({\bf P}^{\prime}_{1,1})^2}{||{\bf P}_{2,1}||^2||{\bf P}_{1,1}||^4} + \frac{1}{2}\frac{{\bf P}_{2,2}({\bf P}_{1,1})^2 \cdot {\bf P}^{\prime}_{2,2}({\bf P}^{\prime}_{1,1})^2}{||{\bf P}_{2,2}||^2||{\bf P}_{1,1}||^4}+\\
        &+\frac{{\bf P}_{3,1}{\bf P}_{1,1} \cdot {\bf P}^{\prime}_{3,1}{\bf P}^{\prime}_{1,1}}{||{\bf P}_{3,1}||^2||{\bf P}_{1,1}||^2}+
        \frac{{\bf P}_{3,2}{\bf P}_{1,1} \cdot {\bf P}^{\prime}_{3,2}{\bf P}^{\prime}_{1,1}}{||{\bf P}_{3,2}||^2||{\bf P}_{1,1}||^2}+
        \frac{{\bf P}_{3,3}{\bf P}_{1,1} \cdot {\bf P}^{\prime}_{3,3}{\bf P}^{\prime}_{1,1}}{||{\bf P}_{3,3}||^2||{\bf P}_{1,1}||^2}+\\
        &+\frac{1}{2}\frac{({\bf P}_{2,1})^2 \cdot ({\bf P}^{\prime}_{2,1})^2}{||{\bf P}_{2,1}||^4}+ \frac{{\bf P}_{2,1}{\bf P}_{2,2} \cdot {\bf P}^{\prime}_{2,1}{\bf P}^{\prime}_{2,2}}{||{\bf P}_{2,1}||^2||{\bf P}_{2,2}||^2}
        +\textcolor{red}{\gamma_1}\cdot\frac{1}{2}\frac{({\bf P}_{2,2})^2 \cdot ({\bf P}^{\prime}_{2,2})^2}{||{\bf P}_{2,2}||^4}+\\
        &+\frac{{\bf P}_{4,1} \cdot {\bf P}^{\prime}_{4,1}}{||{\bf P}_{4,1}||^2} + \frac{{\bf P}_{4,2} \cdot {\bf P}^{\prime}_{4,2}}{||{\bf P}_{4,2}||^2} + 
        \frac{{\bf P}_{4,3} \cdot {\bf P}^{\prime}_{4,3}}{||{\bf P}_{4,3}||^2} + \frac{{\bf P}_{4,4} \cdot {\bf P}^{\prime}_{4,4}}{||{\bf P}_{4,4}||^2} + \\
        &+\textcolor{red}{\gamma_2} \cdot \left( ({\bf P}_{2,2})^2 \cdot {\bf P}^{\prime}_{4,4} + {\bf P}_{4,4} \cdot ({\bf P}^{\prime}_{2,2})^2\right)
    \end{aligned}
    \label{Cauchy 4 level}
\end{align}
where the time-variables from the previous levels have the same modules:
\begin{align}
    \begin{aligned}
        ||{\bf P}_{1,1}||^2 &= 1 &\hspace{15mm}||{\bf P}_{2,1}||^2 &= -\frac{2}{\sigma_2} &\hspace{15mm} ||{\bf P}_{2,2}||^2 &= \frac{2}{\sigma_2} \\
        ||{\bf P}_{3,1}||^2 &= \frac{3}{\sigma_2^2} &\hspace{15mm} ||{\bf P}_{3,2}||^2 &= -\frac{8}{\sigma_2^2} &\hspace{15mm} ||{\bf P}_{3,3}||^2 &= \frac{24}{\sigma _2^2 \left(1-\frac{4 \sigma _2^3}{\sigma _3^2}\right)}
    \end{aligned}
\end{align}
New time-variables have the following modules:
\begin{equation}
    ||{\bf P}_{4,1}||^2 = \frac{4}{\sigma_2^3} \hspace{15mm} ||{\bf P}_{4,2}||^2 = -\frac{20}{\sigma_2^3} \hspace{15mm} ||{\bf P}_{4,3}||^2 = -\frac{144}{\sigma _2^3 \left(1-\frac{4 \sigma _2^3}{\sigma _3^2}\right)}
\end{equation}
\begin{align}
    \begin{aligned}
        ||{\bf P}_{4,4}||^2 &= B^2 \cdot \frac{9 \sigma _2^3 }{5 \sigma _3^8 (Y_{1,2|4}^{2,2})^2} \cdot \Big(12 \sigma _3^4 \sigma _2^6 \left(3993 \tau ^2+968 \tau +282\right)+10 \sigma _3^6 \sigma _2^3 \left(1089 \tau ^2-61\right) -\\
        &-128 \sigma _3^2 \sigma _2^9 (363 \tau +64)+11264 \sigma _2^{12}+25 \sigma _3^8 \Big)
    \end{aligned}
\end{align}
We observe the appearance of anomalous terms:
\begin{equation}
    \textcolor{red}{\gamma_1} = B^2 \cdot \frac{5 \sigma _2^4 }{16 \sigma _3^6} \cdot \left(2 \sigma _3^4 \sigma _2^3 \left(99 \tau ^2+24 \tau +1\right)+45 \sigma _3^6 \tau ^2-12 \sigma _3^2 \sigma _2^6 (16 \tau +3)+112 \sigma _2^9\right)
\end{equation}
\begin{equation}
    \textcolor{red}{\gamma_2} = B^2 \cdot \frac{3 \sigma _2^4 }{2 \sigma _3^7 Y_{1,2|4}^{2,2}} \cdot \left(2 \sigma _3^4 \sigma _2^3 \left(1089 \tau ^2+264 \tau +61\right)+5 \sigma _3^6 \left(99 \tau ^2-1\right)-8 \sigma _3^2 \sigma _2^6 (264 \tau +67)+512 \sigma _2^9\right)
\end{equation}

\bigskip

In the $2d$ limit $h_1 = -1, h_2 = 1, h_3 = 0$ ordinary Schur polynomials are reproduced in the normalization where the coefficient in front of $p_1^4$ is equal to one and identification of time variables ${\bf P}_{a,1} \to p_a$. In our normalization of times the coefficients of higher times ${\bf P}_{a,i > 1}$ are not automatically vanish for all higher times in the $2d$ limit, therefore we set ${\bf P}_{a,i > 1} \to 0$. It is not a problem, all higher times can be rescaled by $\sigma_3$ and the corresponding coefficent in the 3-Schur polynomial will vanish automatically in the $2d$ limit. We keep our convention because of the grading that was explained near \eqref{grading}.
\begin{align}
\begin{aligned}
    {\bf S}_{\, \begin{ytableau}
        \ \\
        \ \\
        \ \\
        \
    \end{ytableau}} \Big|_{\substack{h_1 = -1 \\ h_2 = 1 \\ h_3 = 0}} &= p_1^4-6 p_2 p_1^2+8 p_3 p_1+3 p_2^2-6 p_4, &\hspace{5mm}
    {\bf S}_{\, \begin{ytableau}
        \ & \ & \ & \ 
    \end{ytableau}} \Big|_{\substack{h_1 = -1 \\ h_2 = 1 \\ h_3 = 0}} &= p_1^4+6 p_2 p_1^2+8 p_3 p_1+3 p_2^2+6 p_4 \\
    {\bf S}_{\, \begin{ytableau}
        \ & \ \\
        \ \\
        \
    \end{ytableau}} \Big|_{\substack{h_1 = -1 \\ h_2 = 1 \\ h_3 = 0}} &=p_1^4-2 p_2 p_1^2-p_2^2+2 p_4, &\hspace{5mm}
    {\bf S}_{\, \begin{ytableau}
        \ & \ & \ \\
        \ 
    \end{ytableau}} \Big|_{\substack{h_1 = -1 \\ h_2 = 1 \\ h_3 = 0}} &=p_1^4+2 p_2 p_1^2-p_2^2-2 p_4 \\
    {\bf S}_{\, \begin{ytableau}
        \ & \ \\
        \ & \
    \end{ytableau}} \Big|_{\substack{h_1 = -1 \\ h_2 = 1 \\ h_3 = 0}} &=p_1^4-4 p_3 p_1+3 p_2^2 & &
\end{aligned}
\end{align}

\bigskip
\newpage
\subsection{5 level}
On the level 5 there are 24 basis monomials: $ {\bf P}_{5,1}, \ {\bf P}_{5,2}, \ {\bf P}_{5,3}, \ {\bf P}_{5,4}, \ {\bf P}_{5,5}, \ {\bf P}_{3,1}{\bf P}_{2,1}, \ {\bf P}_{3,2}{\bf P}_{2,1}, \ {\bf P}_{3,3}{\bf P}_{2,1}, \ {\bf P}_{3,1}{\bf P}_{2,2}, $ \\ 
$ \ {\bf P}_{3,2}{\bf P}_{2,2}, \ {\bf P}_{3,3}{\bf P}_{2,2}, \ {\bf P}_{4,1}{\bf P}_{1,1}, \ {\bf P}_{4,2}{\bf P}_{1,1}, \ {\bf P}_{4,3}{\bf P}_{1,1}, \ {\bf P}_{4,4}{\bf P}_{1,1}, \ ({\bf P}_{2,1})^2{\bf P}_{1,1}, \ {\bf P}_{2,1}{\bf P}_{2,2}{\bf P}_{1,1}, \ ({\bf P}_{2,2})^2{\bf P}_{1,1}, {\bf P}_{3,1}({\bf P}_{1,1})^2,$ \\
$\ {\bf P}_{3,2}({\bf P}_{1,1})^2, \ {\bf P}_{3,3}({\bf P}_{1,1})^2, \ {\bf P}_{2,1} ({\bf P}_{1,1})^3, \ {\bf P}_{2,2} ({\bf P}_{1,1})^3, \ ({\bf P}_{1,1})^5$ and $3d$ Young diagrams:
\begin{align}
    \begin{aligned}
        \begin{ytableau}
            \ & \ & \ & \ & \
        \end{ytableau}, \begin{ytableau}
            \ & \ & \ & \ \\
            \
        \end{ytableau},  \begin{ytableau}
            \ & \ & \ \\
            \ & \ 
        \end{ytableau},  \begin{ytableau}
            \ & \ & \ \\
            \ \\
            \
        \end{ytableau}, \begin{ytableau}
            \ & \ \\
            \ & \ \\
            \
        \end{ytableau}, \begin{ytableau}
            \ & \  \\
            \ \\
            \ \\
            \ 
        \end{ytableau}, \begin{ytableau}
            \ \\
            \ \\
            \ \\
            \ \\
            \
        \end{ytableau}, \\
        \begin{ytableau}
            \yt{2} & \ & \ & \ \\
        \end{ytableau}, 
        \begin{ytableau}
            \yt{2} \\
            \ \\
            \ \\
            \
        \end{ytableau}, 
        \begin{ytableau}
            \yt{2} & \yt{2} & \
        \end{ytableau},
        \begin{ytableau}
            \yt{2} \\
            \yt{2} \\
            \
        \end{ytableau},  \begin{ytableau}
            \yt{2} & \yt{2} \\
            \ 
        \end{ytableau},  \begin{ytableau}
            \yt{2} & \ \\
            \yt{2} 
        \end{ytableau}, \begin{ytableau}
            \yt{2} & \ \\
            \ \\
            \
        \end{ytableau}, \begin{ytableau}
            \yt{2} & \ & \ \\
            \ 
        \end{ytableau},  \begin{ytableau}
            \yt{2} & \ \\
            \ & \
        \end{ytableau}, \\
        \begin{ytableau}
            \yt{3} & \ & \  
        \end{ytableau},
        \begin{ytableau}
            \yt{3} \\
            \ \\
            \ 
        \end{ytableau},  \begin{ytableau}
            \yt{3} & \  \\
            \ 
        \end{ytableau},  \begin{ytableau}
            \yt{3} & \yt{2} \\
        \end{ytableau},  \begin{ytableau}
            \yt{3}  \\
            \yt{2} 
        \end{ytableau},
        \begin{ytableau}
            \yt{4} & \ \\
        \end{ytableau},
        \begin{ytableau}
            \yt{4} \\
            \
        \end{ytableau}, \begin{ytableau}
            \yt{5} \\
        \end{ytableau}
    \end{aligned}
\end{align}
As usual, we fix the basis in time-variables ${\bf P}_{5,i}$ by the following conditions:
\begin{equation}
    X_{1,j|k}^{1,b} = Y_{1,j|k}^{1,b} = 0, \hspace{15mm} k \not= i+j-1
\end{equation}
Next we list the Yangian relations and the answers for the ansatz coefficients:
\begin{equation}
    \left( [\psi_2, \psi_3]\right) \cdot {\bf P}_{4,2} = 0
\end{equation}
\begin{align}
\begin{aligned}
     X_{1,1|1}^{1,4} &= -12 \, \sigma_2, &\hspace{15mm} Y_{1,1|1}^{1,4} &= 15, \\
     X_{1,2|2}^{1,4} &= -9 \, \sigma_2, &\hspace{15mm} Y_{1,2|2}^{1,4} &= 18, \\
     X_{1,3|3}^{1,4} &= -6 \, \sigma_2, &\hspace{15mm} Y_{1,3|3}^{1,4} &= 21, \\
     X_{1,4|4}^{1,4} &= -3 \, \sigma_2, &\hspace{15mm} Y_{1,4|4}^{1,4} &= 24. \\
\end{aligned}
\end{align}
We impose the following condition for arbitrary $\alpha_i$:
\begin{equation}
    \left( \psi_3 - [e_2,f_1] \right) \cdot \left( \alpha_1 {\bf P}_{3,1} {\bf P}_{1,1} + \alpha_2 {\bf P}_{3,2} {\bf P}_{1,1}+ \alpha_3 {\bf P}_{3,3} {\bf P}_{1,1} + \alpha_4 {\bf P}_{4,1} +\alpha_5 {\bf P}_{4,2} + \alpha_6 {\bf P}_{4,3} + \alpha_7 {\bf P}_{4,4} \right)  = 0
\end{equation}
\begin{align}
    \begin{aligned}
        X_{1,1|1}^{2,3}&=-18 \, \sigma_2, & \hspace{10mm} X_{1,1|2}^{2,3}&=0,  & \hspace{10mm} X_{1,1|3}^{2,3}&=0, &\hspace{10mm} X_{1,1|4}^{2,3}&=0, \\
        X_{1,2|1}^{2,3}&=0,  & \hspace{10mm} X_{1,2|2}^{2,3}&=- \frac{42}{5} \, \sigma_2,  & \hspace{10mm} X_{1,2|3}^{2,3}&=0, &\hspace{10mm} X_{1,2|4}^{2,3}&= -\frac{198}{5Y_{1,2|4}^{2,2}} \, \sigma_2, \\
        X_{1,3|1}^{2,3}&=0,  & \hspace{10mm} X_{1,3|2}^{2,3}&=0,  & \hspace{10mm} X_{1,3|3}^{2,3}&= - \frac{18}{7} \, \sigma_2, &\hspace{10mm} X_{1,3|4}^{2,3}&=0, \\
        X_{2,1|1}^{2,3}&=0, & \hspace{10mm} X_{2,1|2}^{2,3}&=-\frac{18}{5}\, \sigma_2,  & \hspace{10mm} X_{2,1|3}^{2,3}&=0, &\hspace{10mm} X_{2,1|4}^{2,3}&= -\frac{297}{5 Y_{1,2|4}^{2,2}} \, \sigma_2, \\
        X_{2,2|1}^{2,3}&=0,  & \hspace{10mm} X_{2,2|2}^{2,3}&=0,  & \hspace{10mm} X_{2,2|3}^{2,3}&=\frac{4}{7} \sigma _2 \left(\frac{16 \sigma _2^3}{\sigma _3^2}-33 \tau -4\right), &\hspace{10mm} X_{2,2|4}^{2,3}&=0. \\
    \end{aligned}
\end{align}

\begin{align}
    \begin{aligned}
        Y_{1,1|1}^{2,3}&=15, & \hspace{10mm} Y_{1,1|2}^{2,3}&=0,  & \hspace{10mm} Y_{1,1|3}^{2,3}&=0, &\hspace{10mm} Y_{1,1|4}^{2,3}&=0, \\
        Y_{1,2|1}^{2,3}&=0,  & \hspace{10mm} Y_{1,2|2}^{2,3}&=21,  & \hspace{10mm} Y_{1,2|3}^{2,3}&=0, &\hspace{10mm} Y_{1,2|4}^{2,3}&= Y_{1,2|4}^{2,2}, \\
        Y_{1,3|1}^{2,3}&=0,  & \hspace{10mm} Y_{1,3|2}^{2,3}&=0,  & \hspace{10mm} Y_{1,3|3}^{2,3}&= 27, &\hspace{10mm} Y_{1,3|4}^{2,3}&=0, \\
        Y_{2,1|1}^{2,3}&=0, & \hspace{10mm} Y_{2,1|2}^{2,3}&=24,  & \hspace{10mm} Y_{2,1|3}^{2,3}&=0, &\hspace{10mm} Y_{2,1|4}^{2,3}&= 4 Y_{1,2|4}^{2,2}, \\
        Y_{2,2|1}^{2,3}&=0,  & \hspace{10mm} Y_{2,2|2}^{2,3}&=0,  & \hspace{10mm} Y_{2,2|3}^{2,3}&= B \cdot 90 \left(12 \tau +1-\frac{14 \sigma _2^3}{\sigma _3^2}\right), &\hspace{10mm} Y_{2,2|4}^{2,3}&=0. \\
    \end{aligned}
\end{align}

\begin{align}
    \begin{aligned}
        U_{1,1}^{5} &= -70 \, \frac{\sigma_3}{\sigma_2}, &\hspace{10mm} U_{1,2}^{5} &= 120 \, \frac{\sigma_3}{\sigma_2}, &\hspace{10mm} U_{1,3}^{5} &= 0, &\hspace{10mm} U_{1,4}^{5} &= 0, \\
        U_{2,1}^{5} &= -15 \, \frac{\sigma_3}{\sigma_2}, &\hspace{10mm} U_{2,2}^{5} &= 5 \, \frac{\sigma_3}{\sigma_2}, &\hspace{10mm} U_{2,3}^{5} &= 63 \, \frac{\sigma_3}{\sigma_2}, &\hspace{10mm} U_{2,4}^{5} &= 0, \\
        U_{3,1}^{5} &= 0, &\hspace{10mm} U_{3,2}^{5} &= 5 \, \frac{\sigma_3}{\sigma_2}\, \left( 4\, \frac{\sigma_2^3}{\sigma_3^2} - 1 \right), &\hspace{10mm} U_{3,3}^{5} &= 5 \, \frac{\sigma_3}{\sigma_2}, &\hspace{10mm} U_{3,4}^{5} &= \frac{20}{7}\,\tau \cdot Y_{1,2|4}^{2,2} \cdot \frac{\sigma_3}{\sigma_2}, \\
        U_{4,1}^{5} &= 0, &\hspace{10mm} U_{4,2}^{5} &= 0, &\hspace{10mm} U_{4,3}^{5} &= \frac{5}{4} \cdot U_{4,3}^{4}, &\hspace{10mm} U_{4,4}^{5} &= 5 \, \frac{\sigma_3}{\sigma_2}. \\
    \end{aligned}
\end{align}
\begin{align}
    \begin{aligned}
         \text{YRel}_{0,0} \cdot \left( \alpha_1 ({\bf P}_{1,1})^3 + \alpha_2 {\bf P}_{2,1}{\bf P}_{1,1} + \alpha_3 {\bf P}_{2,2}{\bf P}_{1,1} + \alpha_4 {\bf P}_{3,1}  + \alpha_5 {\bf P}_{3,2}  + \alpha_6 {\bf P}_{3,3} \right) & =0 \\
         \text{YRel}_{1,1} \cdot \left( \alpha_1 ({\bf P}_{1,1})^3 + \alpha_2 {\bf P}_{2,1}{\bf P}_{1,1} + \alpha_3 {\bf P}_{2,2}{\bf P}_{1,1} + \alpha_4 {\bf P}_{3,1}  + \alpha_5 {\bf P}_{3,2}  + \alpha_6 {\bf P}_{3,3} \right) & =0 \\
    \end{aligned}
\end{align}
\begin{align}
    \begin{aligned}
        X_{1,1|5}^{2,3}&=0, & \hspace{10mm} X_{1,2|5}^{2,3}&=0, & \hspace{10mm} X_{1,3|5}^{2,3}&=-\frac{1026}{7Y_{1,3|5}^{2,3}} \, \sigma_2, \\
        X_{2,1|5}^{2,3}&=0, & \hspace{10mm} X_{2,2|5}^{2,3}&=-\frac{468 \left(\sigma _2 \sigma _3^2 (33 \tau +4)-16 \sigma _2^4\right)}{35 \sigma _3^2 Y_{1,3|5}^{2,3}}, & \hspace{10mm} X_{2,3|5}^{2,3}&=0. \\
    \end{aligned}
\end{align}
\begin{align}
    \begin{aligned}
        Y_{1,1|5}^{2,3}&=0, & \hspace{10mm} Y_{1,2|5}^{2,3}&=0, & \hspace{10mm} & \\
        Y_{2,1|5}^{2,3}&=0, & \hspace{10mm} Y_{2,2|5}^{2,3}&=\frac{26 Y_{1,3|5}^{2,3} \left(\sigma _3^2 (12 \tau +1)-14 \sigma _2^3\right)}{19 \sigma _3^2}, & \hspace{10mm} Y_{2,3|5}^{2,3}&=0. 
    \end{aligned}
\end{align}
\begin{align}
    \begin{aligned}
        U_{1,5}^{5} &= 0, &\hspace{10mm} U_{2,5}^{5} &= 0, &\hspace{10mm} U_{3,5}^{5} &= 0, \\  U_{5,1}^{5} &= 0, &\hspace{10mm} U_{5,2}^{5} &= 0, &\hspace{10mm} U_{5,3}^{5} &= 0.
    \end{aligned}
\end{align}
\begin{align}
    \begin{aligned}
        U_{5,5}^{5} &= 5 \, \frac{\sigma_3}{\sigma_2}, \\
        U_{5,4}^{5} &= \frac{3 Y_{1,2|4}^{2,2} \left(\sigma _3^2 (156 \tau -7)+28 \sigma _2^3\right)}{7 \sigma _2 \sigma _3 Y_{1,3|5}^{2,3}} \\
        U_{4,5}^{5} &=-\frac{3 Y_{1,3|5}^{2,3} \left(-2 \sigma _3^4 \sigma _2^3 \left(15444 \tau ^2+7059 \tau +56\right)+8 \sigma _3^2 \sigma _2^6 (1179 \tau -28)+375 \sigma _3^6 \tau +2688 \sigma _2^9\right)}{\sigma _2 \sigma _3^3 Y_{1,2|4}^{2,2} \left(\sigma _3^2 (156 \tau -7)+28
   \sigma _2^3\right)}\\
    \end{aligned}
\end{align}
Now provide explicit formulas for 3-Schur polynomials using the following notation:
\begin{equation}
    C = \frac{\sigma _3^4}{1824 \sigma _2^6-2 \sigma _3^2 \sigma _2^3 (1881 \tau +592)+\sigma _3^4 (52-231 \tau )}
\end{equation}
\begin{equation}
    R(h_i)=\frac{5 \left(-2 \sigma _3^2 \sigma _2^3 (99 \tau +16) h_i^3+6 \sigma _3^3 \sigma _2^2 (66 \tau +53) h_i^2+\sigma _3^4 \sigma _2 (33 \tau +74) h_i+624 \sigma _2^5 h_i^5+516 \sigma _3 \sigma _2^4 h_i^4+\sigma _3^5 (52-231 \tau )\right)}{2 \sigma _3^3 h_i^5}
\end{equation}
\begin{align}
    \begin{aligned}
        T(h_i) =& -\frac{855}{7 \sigma _3^4 Y_{1,3|5}^{2,3} h_i^7} \Big(22 \sigma _3^2 \sigma _2^5 (9 \tau -4) h_i^5-2 \sigma _3^3 \sigma _2^4 (198 \tau +59) h_i^4+\sigma _3^4 \sigma _2^3 (26-33 \tau ) h_i^3+\sigma _3^5 \sigma _2^2 (231 \tau +43) h_i^2+96 \sigma _2^7 h_i^7- \\ 
    -&96 \sigma _3 \sigma _2^6 h_i^6+20 \sigma
   _3^6 \sigma _2 h_i+5 \sigma _3^7\Big)
    \end{aligned}
\end{align}

\begin{align}
    \begin{aligned}
        \lambda_{\,\begin{ytableau}
            \ \\
            \ \\
            \ \\
            \ \\
            \ \\
        \end{ytableau}} &= 60 h_1 - 10 \frac{\sigma_3}{\sigma_2} \\
        {\bf S}_{\, \begin{ytableau}
            \ \\
            \ \\
            \ \\
            \ \\
            \ \\
        \end{ytableau}} &= {\bf P}_{1,1}^5 + 10 h_1 \cdot {\bf P}_{1,1}^3 {\bf P}_{2,1} + 10 \frac{\sigma_3}{h_1^2} \cdot {\bf P}_{1,1}^3 {\bf P}_{2,2} + 20 h_1^2 \cdot {\bf P}_{1,1}^2 {\bf P}_{3,1} + 15 \frac{\sigma_3}{h_1} \cdot {\bf P}_{1,1}^2 {\bf P}_{3,2} + \frac{5}{2} K(h_1) \cdot {\bf P}_{1,1}^2 {\bf P}_{3,3} +  \\
        &+30 h_1^3 \cdot {\bf P}_{1,1} {\bf P}_{4,1} + 18 \sigma_3 \cdot {\bf P}_{1,1} {\bf P}_{4,2} + \frac{5}{2} h_1 K(h_1) \cdot {\bf P}_{1,1} {\bf P}_{4,3} + 5 M(h_1) \cdot {\bf P}_{1,1} {\bf P}_{4,4} + 15 h_1^2 \cdot {\bf P}_{1,1} {\bf P}_{2,1}^2 + \\
        &+30 \frac{\sigma_3}{h_1} \cdot {\bf P}_{1,1} {\bf P}_{2,1}{\bf P}_{2,2} + 5 L(h_1) \cdot {\bf P}_{1,1} {\bf P}_{2,2}^2 +  20 h_1^3 \cdot {\bf P}_{2,1} {\bf P}_{3,1} + 15 \sigma_3 \cdot {\bf P}_{2,1} {\bf P}_{3,2} + \frac{5}{2} h_1 K(h_1) \cdot {\bf P}_{2,1} {\bf P}_{3,3} + \\
        &+20 \sigma_3 \cdot {\bf P}_{2,2} {\bf P}_{3,1} + 5 h_1 L(h_1) B \cdot {\bf P}_{2,2} {\bf P}_{3,2} + K(h_1) R(h_1) C \cdot {\bf P}_{2,2} {\bf P}_{3,3} + 24 h_1^4 \cdot {\bf P}_{5,1} + 12 h_1 \sigma_3 \cdot {\bf P}_{5,2} +\\
        &+\frac{10}{7} h_1^2 K(h_1) \cdot {\bf P}_{5,3} + \frac{5}{2} h_1 M(h_1) \cdot {\bf P}_{5,4} + K(h_1) T(h_1) \cdot {\bf P}_{5,5}
    \end{aligned}
\end{align}

\begin{align}
    \begin{aligned}
        \lambda_{\,\begin{ytableau}
            \ & \ \\
            \ \\
            \ \\
            \ 
        \end{ytableau}} &= 36 h_1+6h_2 - 10 \frac{\sigma_3}{\sigma_2} \\
        {\bf S}_{\, \begin{ytableau}
            \ & \ \\
            \ \\
            \ \\
            \ 
        \end{ytableau}} &=  {\bf P}_{1,1}^5 + (6h_1+h_2) \cdot {\bf P}_{1,1}^3 {\bf P}_{2,1} + 
        \frac{h_3^2}{\sigma_3} \left(\left(6 h_1+h_2\right) h_3-10 \sigma _2\right)\cdot {\bf P}_{1,1}^3 {\bf P}_{2,2}+
        h_1 \left(6 h_1+h_2-2 h_3\right) \cdot {\bf P}_{1,1}^2 {\bf P}_{3,1}+\\
        &+\frac{3 h_3^2}{4 \sigma _3} \left(7 h_3 \sigma _2+2 h_2 \left(4 h_3^2+5 \sigma _2\right)+3h_3^3\right) \cdot {\bf P}_{1,1}^2 {\bf P}_{3,2}
        -\frac{8 h_3^2+5 h_2 h_3+20 \sigma _2}{8 h_2^2} K(h_1) \cdot {\bf P}_{1,1}^2 {\bf P}_{3,3}-6 h_1^2 h_3 \cdot {\bf P}_{1,1} {\bf P}_{4,1}+\\
        &+\frac{3 h_3}{5 h_2} \left(5 \left(h_2+2 h_3\right) \sigma _2-6 h_1 h_3^2\right) \cdot {\bf P}_{1,1} {\bf P}_{4,2}+
        \frac{5 h_3 \sigma _2+h_2^3+6 h_1 h_2^2-4 h_3^2 h_2+h_3^3}{4h_2^2} K(h_1) \cdot {\bf P}_{1,1} {\bf P}_{4,3}+\\
        & -\frac{h_3^2+5 \sigma _2}{h_2^2} B \cdot M(h_1) \cdot {\bf P}_{1,1} {\bf P}_{4,4}
        -3 h_1 h_3 \cdot {\bf P}_{1,1} {\bf P}_{2,1}^2+
        \frac{3 h_3^2 \left(5 h_2+6 h_3\right) \sigma _2-6 h_1
   h_3^4}{\sigma _3} \cdot {\bf P}_{1,1} {\bf P}_{2,1}{\bf P}_{2,2}-\\
       & -\frac{h_3^2+5 \sigma _2}{h_2^2}  B \cdot L(h_1) \cdot {\bf P}_{1,1} {\bf P}_{2,2}^2+
       5 h_1^2 h_2 \cdot {\bf P}_{2,1} {\bf P}_{3,1} + 
       \frac{3 h_3^2 \left(5 h_3^2+9 \sigma _2\right)}{4 h_2} \cdot {\bf P}_{2,1} {\bf P}_{3,2} + \\
       &+ \frac{27 h_3 \sigma _2-10 h_2^3+15 h_3^3}{8 h_2^2} K(h_1) \cdot {\bf P}_{2,1} {\bf P}_{3,3} + 
        h_3 \left(-\frac{8 h_3^3}{h_2}+13 h_1 h_3+5 \sigma _2\right) \cdot {\bf P}_{2,2} {\bf P}_{3,1} + \\
       &+\frac{27 h_3 \sigma _2-10 h_2^3+15 h_3^3}{4 h_2^2} B \cdot L(h_1) \cdot {\bf P}_{2,2} {\bf P}_{3,2} +\frac{B \cdot K(h_1)}{8 h_1^6 h_2^2 \sigma _3^3} \Big(-4 h_1^4 \sigma _3^2 \sigma _2^4 (42 \tau -5)+2 h_1^3
   \sigma _3^3 \sigma _2^3 (111 \tau +8)+ \\
   &+6 h_1^2 \sigma _3^4 \sigma _2^2 (8 \tau -1)-3 h_1 \sigma _3^5 \sigma _2 (59 \tau +2)-16 h_1^6 \sigma _2^6+48 h_1^5 \sigma _3 \sigma _2^5+75\sigma _3^6 \tau \Big) \cdot {\bf P}_{2,2} {\bf P}_{3,3} + 
      6 h_1^3 h_2 \cdot {\bf P}_{5,1} + \\
       &+\frac{3 \sigma _3 \left(14 h_3 \sigma _2-5 h_2 \sigma _2+10
   h_3^3\right)}{10 h_2^2} \cdot {\bf P}_{5,2} + 
       \frac{h_1 \left(22 h_3 \sigma _2+5 h_2 \sigma _2+10
   h_3^3\right)}{28 h_2^2} K(h_1) \cdot {\bf P}_{5,3} + \\
       &+\frac{17 h_3 \sigma _2+10 h_2 \sigma _2+5 h_3^3}{8 h_2^2} B \cdot M(h_1) \cdot {\bf P}_{5,4} + 
       -\frac{9 B \cdot K(h_1)}{28 h_1^8 h_2^2 \sigma
   _3^4 Y_{1,3|5}^{2,3}} \Big(24 h_1^6 \sigma _3^2 \sigma _2^6 (55 \tau -17)-\\
   &-2h_1^5 \sigma _3^3 \sigma _2^5 (1485 \tau +124)+2 h_1^4 \sigma
   _3^4 \sigma _2^4 (168 \tau +43)+9 h_1^3 \sigma _3^5 \sigma
   _2^3 (235 \tau +22)-3 h_1^2 \sigma _3^6 \sigma _2^2 (319 \tau
   +7)+640 h_1^8 \sigma _2^8-\\
   &-800 h_1^7 \sigma _3 \sigma
   _2^7+\sigma _3^8 (156 \tau -7)\Big) \cdot {\bf P}_{5,5} 
    \end{aligned}
\end{align}
\begin{align}
    \begin{aligned}
        \lambda_{\,\begin{ytableau}
            \yt{2} & \ \\
            \ & \ 
        \end{ytableau}} &= -6h_3 - 10 \frac{\sigma_3}{\sigma_2} \\
        {\bf S}_{\, \begin{ytableau}
            \yt{2} & \ \\
            \ & \
        \end{ytableau}} &= {\bf P}_{1,1}^5 + 
        (-h_3) \cdot {\bf P}_{1,1}^3 {\bf P}_{2,1}
        -\frac{3 h_3^2 \sigma _2+h_3^4-4 \sigma _2^2}{\sigma_3}\cdot {\bf P}_{1,1}^3 {\bf P}_{2,2}
       -\left(h_1-h_2\right)^2 \cdot {\bf P}_{1,1}^2 {\bf P}_{3,1}+\\
        &+\frac{11 h_3^2 \sigma _2+9 h_3^4-4 \sigma _2^2}{4 h_1h_2} \cdot {\bf P}_{1,1}^2 {\bf P}_{3,2}+
        +\frac{3 \left(h_2 h_3-2 \sigma _2\right) \left(h_3^2-2
   \sigma _2\right)}{8 h_2^2 h_3^2} K(h_1) \cdot {\bf P}_{1,1}^2 {\bf P}_{3,3}+
        2 \sigma _3 \cdot {\bf P}_{1,1} {\bf P}_{4,1}+\\
        &+\frac{3 \left(9 h_3^4 \sigma _2+7 h_3^2 \sigma _2^2+4
   h_3^6+4 \sigma _2^3\right)}{10 \sigma _3} \cdot {\bf P}_{1,1} {\bf P}_{4,2}+
        \frac{\left(h_2 h_3-2 \sigma _2\right) \left(4 h_3^2+7
   \sigma _2\right)}{24 h_2^2 h_3} K(h_1) \cdot {\bf P}_{1,1} {\bf P}_{4,3}-\\
        &-\frac{6 B}{5 \sigma _3^5
   Y_{1,2|4}^{2,2}} \Big(30 h_3^{14} \sigma _2^2 (33 \tau
   -7)+h_3^{12} \sigma _2^3 (4158 \tau -61)+2 h_3^{10}
   \sigma _2^4 (4752 \tau +551)+9 h_3^8 \sigma _2^5
   (1672 \tau +291)+\\
   &+6 h_3^6 \sigma _2^6 (2079 \tau
   +251)+h_3^4 \sigma _2^7 (2970 \tau -2291)-12 h_3^2
   \sigma _2^8 (66 \tau +247)-75 h_3^{16} \sigma _2-10
   h_3^{18}+384 \sigma _2^9\Big) \cdot {\bf P}_{1,1} {\bf P}_{4,4}+\\
    &+(-3 h_1 h_2) \cdot {\bf P}_{1,1} {\bf P}_{2,1}^2
       -(6 h_3^2+4 \sigma_2) \cdot {\bf P}_{1,1} {\bf P}_{2,1}{\bf P}_{2,2}
       -\frac{B}{\sigma _3^4}\Big(h_3^{12} \sigma _2 (189 \tau +2)+h_3^{10}
   \sigma _2^2 (432 \tau +11)+ \\
   &+6 h_3^8 \sigma _2^3 (114
   \tau +7)+9 h_3^6 \sigma _2^4 (63 \tau +2)+h_3^4
   \sigma _2^5 (135 \tau -68)-h_3^2 \sigma _2^6 (36
   \tau +137)+45 h_3^{14} \tau +132 \sigma
   _2^7\Big) \cdot {\bf P}_{1,1} {\bf P}_{2,2}^2-\\
    &-h_3 \left(h_3^2+2 \sigma _2\right) \cdot {\bf P}_{2,1} {\bf P}_{3,1}
    -\Big(\frac{5 h_3 \sigma _2}{4}+\frac{3 \sigma_2^2}{h_3}+\frac{3 h_3^3}{4}\Big) \cdot {\bf P}_{2,1} {\bf P}_{3,2}+
        \frac{1}{8} h_1 \left(\frac{2 \sigma _2}{h_2
   h_3}-1\right) K(h_1) \cdot {\bf P}_{2,1} {\bf P}_{3,3}- \\
       &-\frac{h_1 h_2 \left(h_3^2-4 \sigma _2\right)}{h_3} \cdot {\bf P}_{2,2} {\bf P}_{3,1}
        -\frac{\left(h_3^3-h_3 \sigma _2\right)B}{4 \sigma _3^4} \Big(5
   h_3^{10} \sigma _2 (18 \tau -1)-2 h_3^8 \sigma _2^2
   (27 \tau +11)-2 h_3^6 \sigma _2^3 (117 \tau
   +11)+\\
   &+h_3^4 \sigma _2^4 (62-171 \tau )+h_3^2 \sigma
   _2^5 (175-36 \tau )+45 h_3^{12} \tau +132 \sigma
   _2^6 \Big) \cdot {\bf P}_{2,2} {\bf P}_{3,2}+
   (-h_3 \sigma_3) \cdot {\bf P}_{5,1}-\\
    &-\frac{h_1^2 \left(h_2 h_3-2 \sigma _2\right)B \cdot K(h_1)}{8 \sigma _3^5}
   \Big(h_3^{10} \sigma _2 (18 \tau -1)+2 h_3^8
   \sigma _2^2 (3 \tau -1)+6 h_3^6 \sigma _2^3 (7 \tau
   +1)+9 h_3^4 \sigma _2^4 (7 \tau +2)+\\
   &+h_3^2 \sigma_2^5 (24 \tau +7)+15 h_3^{12} \tau -28 \sigma
   _2^6\Big) \cdot {\bf P}_{2,2} {\bf P}_{3,3}
     -\frac{19 h_3^4 \sigma _2+7 h_3^2 \sigma _2^2+10
   h_3^6+4 \sigma _2^3}{20 h_1 h_2}\cdot {\bf P}_{5,2}-\\
    &-\frac{\left(h_2 h_3-2 \sigma _2\right) \left(10
   h_3^2+7 \sigma _2\right)}{168 h_2^2} K(h_1) \cdot {\bf P}_{5,3}
    -\frac{3 h_3 \left(h_3^2-\sigma _2\right) B }{20 \sigma _3^5
   Y_{1,2|4}^{2,2}} \Big(5 h_3^{12} \sigma _2^2 (198 \tau +95)+20 h_3^{10}
   \sigma _2^3 (99 \tau +13)-\\
   &-11 h_3^8 \sigma _2^4 (108
   \tau +139)-4 h_3^6 \sigma _2^5 (1287 \tau +956)-11
   h_3^4 \sigma _2^6 (342 \tau +241)+12 h_3^2 \sigma
   _2^7 (25-66 \tau )+160 h_3^{14} \sigma _2+\\
   &+25h_3^{16}+384 \sigma _2^8\Big) \cdot {\bf P}_{5,4}+
    \frac{3 h_1^2 \left(h_2 h_3-2 \sigma_2\right) B \cdot K(h_1)}{56 \sigma _3^6 Y_{1,3|5}^{2,3}} \Big(4h_3^{14} \sigma _2 (156 \tau -7)-h_3^{12} \sigma
   _2^2 (618 \tau +49)-\\
   &-4 h_3^{10} \sigma _2^3 (903
   \tau -53)+h_3^8 \sigma _2^4 (487-5208 \tau )-4
   h_3^6 \sigma _2^5 (2445 \tau +101)-h_3^4 \sigma
   _2^6 (12642 \tau +2255)-\\
   &-28 h_3^2 \sigma _2^7 (198
   \tau +23)+h_3^{16} (156 \tau -7)+2688 \sigma
   _2^8\Big) \cdot {\bf P}_{5,5}
    \end{aligned}
\end{align}

We do not provide here all the 3-Schur polynomials of 5-th level due to the length of the formulas. 
The characteristic equation has a nice form, that is in full agreement with \eqref{eigenvalues}:
\begin{align}
    \boxed{
    \begin{aligned}
        0 &= \det( \psi_3 - 6 u + 4 \frac{\sigma_3}{\sigma_2} ) \ \ \varpropto \ \ \left(u^3+100 \sigma _2 u -1000 \sigma _3\right) \left(u^3+4 \sigma _2 u -8 \sigma _3\right) \left(u^3+\sigma _2 u + \sigma _3\right) \times \\
        &\times\left(u^3+9 \sigma _2 u + 27 \sigma _3\right) \left(u^6+24 \sigma _2 u^4+144 \sigma _2^2 u^2 + 256 \sigma _2^3+1728 \sigma _3^2\right) \times \\
   & \times \left(u^6+62 \sigma _2 u^4-308 \sigma _3 u^3+961 \sigma _2^2 u^2-9548 \sigma _2 \sigma _3 u + 900 \sigma _2^3+29791 \sigma _3^2\right) =\\
   \\
   &=\left(u-10h_1\right) \left(u-10 h_2\right)\left(u-10 h_3\right)\left(u- 2h_1\right) \left(u- 2h_2\right) \left(u- 2h_3\right) \times \\
   &\times \left(u+h_1\right) \left(u+h_2\right) \left(u+h_3\right) \left(u+3h_1\right) \left(u+3h_2\right) \left(u+3h_3\right) \times \\
   & \times \left(u-2h_1-4 h_2\right)\left(u-2h_1-4 h_3\right) \left(u-2h_2-4 h_1\right) \left(u-2h_2-4 h_3\right) \left(u-2h_3-4 h_1\right) \left(u-2h_3-4 h_2\right) \times \\
   &\times \left(u-h_1-6 h_2\right) \left(u-h_1-6 h_3\right) \left(u-h_2-6 h_1\right) \left(u-h_2-6 h_3\right) \left(u-h_3-6 h_1\right) \left(u-h_3-6 h_2\right) 
    \end{aligned}
    }
    \label{char eq 5}
\end{align}
All the 3-Schur polynomials on the 5-th level have correct $2d$ limit, in particular:
\begin{align}
    \begin{aligned}
         {\bf S}_{\, \begin{ytableau}
        \ \\
        \ \\
        \ \\
        \ \\
        \ 
    \end{ytableau}} \Big|_{\substack{h_1 = -1 \\ h_2 = 1 \\ h_3 = 0}} &= p_1^5-10 p_2 p_1^3+20 p_3 p_1^2+15 p_2^2 p_1-30 p_4 p_1-20 p_2 p_3+24 p_5 \\
    {\bf S}_{\, \begin{ytableau}
        \ & \ \\
        \ \\
        \ \\
        \ 
    \end{ytableau}} \Big|_{\substack{h_1 = -1 \\ h_2 = 1 \\ h_3 = 0}} &= p_1^5-5 p_2 p_1^3+5 p_3 p_1^2+5 p_2 p_3-6 p_5 \\
    \end{aligned}
\end{align}
We checked Cauchy identity and observe the following deviation from the classical formula \eqref{Cauchy id}:
\begin{align}
    \begin{aligned}
        &\sum_{|\pi| = 5} \frac{{\bf S}_{\pi}({\bf P}) \cdot {\bf S}_{\pi}({\bf P}^{\prime})}{|| {\bf S}_{\pi} ||^2} = \exp \left\{ \sum_{a \geqslant i} \frac{{\bf P}_{a,i} \cdot {\bf P}^{\prime}_{a,i}}{|| {\bf P}_{a,i}||^2} \right\} \Big|_{\text{5-th level}} + \textcolor{red}{\gamma_3} \cdot ({\bf P}_{2,2})^2 {\bf P}_{1,1} \cdot ({\bf P}_{2,2}^{\prime})^2 {\bf P}^{\prime}_{1,1} + \\
        &+\textcolor{red}{\gamma_4} \cdot {\bf P}_{4,4} {\bf P}_{1,1} \cdot {\bf P}_{4,4}^{\prime} {\bf P}^{\prime}_{1,1} + \textcolor{red}{\gamma_5} \cdot {\bf P}_{2,2} {\bf P}_{3,2} \cdot {\bf P}_{2,2}^{\prime} {\bf P}^{\prime}_{3,2} + \textcolor{red}{\gamma_6} \cdot {\bf P}_{2,2} {\bf P}_{3,3} \cdot {\bf P}_{2,2}^{\prime} {\bf P}^{\prime}_{3,3} + \\
        &+ \textcolor{red}{\gamma_7} \cdot \left( ({\bf P}_{2,2})^2 {\bf P}_{1,1} \cdot {\bf P}_{4,4}^{\prime} {\bf P}^{\prime}_{1,1} + {\bf P}_{4,4} {\bf P}_{1,1} \cdot ({\bf P}_{2,2}^{\prime})^2 {\bf P}^{\prime}_{1,1} \right)+\\
        &+ \textcolor{red}{\gamma_8} \cdot \left( {\bf P}_{2,2} {\bf P}_{3,2} \cdot {\bf P}_{5,4}^{\prime}  + {\bf P}_{5,4} \cdot {\bf P}_{2,2}^{\prime} {\bf P}^{\prime}_{3,2} \right)+\textcolor{red}{\gamma_9} \cdot \left( {\bf P}_{2,2} {\bf P}_{3,3} \cdot {\bf P}_{5,5}^{\prime}  + {\bf P}_{5,5} \cdot {\bf P}_{2,2}^{\prime} {\bf P}^{\prime}_{3,3} \right)
    \end{aligned}
    \label{Cauchy 5 level}
\end{align}
We omit explicit formulas for the deviation coefficients $\textcolor{red}{\gamma_i}$. Note that the deviation from the 4-th level in the sector of times ${\bf P}_{2,2}^2$, ${\bf P}_{4,4}$ is inherited and new deviation in sectors ${\bf P}_{2,2} {\bf P}_{3,2}$, ${\bf P}_{5,4}$ and ${\bf P}_{2,2} {\bf P}_{3,3}$, ${\bf P}_{5,5}$ occur.

\bigskip

\section{Conclusion
\label{conc}}
 
In this paper we continued \cite{Morozov:2018fga, MThunt} to use the cut-and-join operator approach \cite{MMN}
to attack the problem of $3$-Schur functions \cite{Morozov:2018fjb, Wang1}. This time we applied more knowledge from the Yangian theory \cite{Tsymbaliuk:2014fvq, Maulik:2012wi,Prochazka:2015deb}, where the operator of our interest is identified with $\psi_3$ and the $3$-Schur problem reduces to a search for
the {\it triangular} representation of Yangian $Y(\hat{ \mathfrak{gl}}_1)$ algebra by the differential operators in appropriate set of time-variables ${\bf P}_{a,i}$ with $1\leqslant i\leqslant a$.
 
We present explicit formulas for the triangular representation up to level 5, i.e. the representation on finite set of time-variables ${\bf P}_{a,i}$, where $a = 1,\ldots, 5$ and $i \leqslant a$. While constructing the representation on the level $n$ we choose the basis in the subspace of times ${\bf P}_{n,i}$ using the rules $Y_{1,j|k}^{1,b} \sim \delta_{j,k}$ and $X_{1,j|k}^{1,b} \sim \delta_{j,k}$. 3-Schur polynomials are then computed as the common eigenfunctions of the commuting family of $\psi_n$ operators of the Yangian. 

According to the results of the calculation of 3-Schur polynomials one can see that the choice of basis in times turns out to be unsuccessful. Firstly, the explicit formulas for 3-Schur polynomials are ugly and enormous for 4-th and 5-th levels. Secondly, the deviations from the Cauchy identity on 4,5 levels \eqref{Cauchy 4 level},\eqref{Cauchy 5 level} indicates the loss of orthogonality property. Namely, the anomalous terms are observed in the following sectors: on 4-th level $\left( {\bf P}_{2,2}^2, {\bf P}_{4,4} \right) $; on 5-th level $\left( {\bf P}_{2,2} {\bf P}_{3,2}, {\bf P}_{5,4} \right) $ and $\left( {\bf P}_{2,2} {\bf P}_{3,3}, {\bf P}_{5,3} \right)$. 

On the other hand, the truly {\it invariant} quantities (that do not depend on the choice of basis) -- the characteristic polynomials \eqref{char eq 2}, \eqref{char eq 3}, \eqref{char eq 4}, \eqref{char eq 5} have nice and consize form, with the eigenvalues that obey \eqref{eigenvalues}. In addition, our formula for eigenvalues coincides with the other results presented in the literature \cite{Prochazka:2015deb}. We consider these facts as the indication that our algorithm works properly and the triangular representation of the Yangian is correct.

Our results for 3-Schur polynomials reproduce the previously known $3$-Schur functions of \cite{Wang1, Wang4} at the levels $1-3$, but deviates from those
from \cite{Wang6} at level $4$ -- the first level where
the truly three-dimensional plane partitions appear,
and the problem becomes truly difficult. Presumably, this difference can be eliminated by the change of basis of time-variables.

The idea of general approach seems to be more or less clear now,
still corrections are needed to handle the higher levels
and to find the generic expressions, we postpone this for the future work.
 
Despite the fact that we failed to choose the right basis in the space of time-variables, we managed to develop an algorithm to construct the Yangian representation and compute 3-Schur functions. One should now look for the ways to bypass mounting problems,
for example, by considering alternative implications of Yangian
representation theory and checking if they can be lifted to
the $3d$ representation. A promising direction is to test the commutative subalgebras
and generalized integrable systems, introduced in \cite{MMintsystWLZZ}.

\section*{Acknowledgements}
We are indebted for interesting discussions to S. Barakin, K. Gubarev, K. Khmelevsky, N. Kolganov, E. Lanina, A. Mironov, V. Mishnyakov, E. Musaev, P. Suprun, M. Tsarkov and especially for numerous explanations to D. Galakhov. \\
Our work is supported by the Russian Science Foundation (Grant No.20-71-10073).

\printbibliography
\end{document}